\documentclass[a4paper,11pt]{article}
\pdfoutput=1 % if your are submitting a pdflatex (i.e. if you have
\usepackage{physics}
\usepackage{jheppub} % for details on the use of the package, please
\usepackage[T1]{fontenc} % if needed
\usepackage[dvipsnames]{xcolor}
\usepackage{braket}
\usepackage{bbm}
\usepackage[normalem]{ulem}
\usepackage{hyperref}

\usepackage{import}

\usepackage{tikz}
\usepackage{pgfplots}
\pgfplotsset{compat=newest}
\usepgfplotslibrary{fillbetween}
\usepackage{subcaption}
\usepackage{multirow}

\begin{document}

\title{\boldmath Entanglement asymmetry in the critical XXZ spin chain}

\vspace{.5cm}

\author{Marco Lastres$^{1,2}$, Sara Murciano$^{3,4}$, Filiberto Ares$^{5}$, Pasquale Calabrese$^{5,6}$}
\affiliation{$^{1}$Dipartimento di Fisica dell'Universit\`a di Pisa and Scuola Normale Superiore, I-56127 Pisa, Italy}
\affiliation{$^{2}$Department of Physics, Technical University of Munich, 85748 Garching, Germany}
\affiliation{$^{3}$Walter Burke Institute for Theoretical Physics, Caltech, Pasadena, CA 91125, USA}
\affiliation{$^{4}$Department of Physics and IQIM, Caltech, Pasadena, CA 91125, USA}
\affiliation{$^{5}$SISSA and INFN Sezione di Trieste, via Bonomea 265, 34136 Trieste, Italy.}
\affiliation{$^{6}$International Centre for Theoretical Physics (ICTP), Strada Costiera 11, 34151 Trieste, Italy.}

\abstract{
We study the explicit breaking of a $SU(2)$ symmetry to a $U(1)$ subgroup employing the entanglement asymmetry, a recently introduced observable that measures how much symmetries are broken in a part of extended quantum systems. We consider as specific model the critical XXZ spin chain, which breaks the $SU(2)$ symmetry of spin rotations except at the isotropic point, and is described by the massless compact boson in the continuum limit. We examine the $U(1)$ subgroup of $SU(2)$ that is broken outside the isotropic point by applying conformal perturbation theory, which we complement with numerical simulations on the lattice. We also analyse the entanglement asymmetry of the full $SU(2)$ group. By relying on very generic scaling arguments, we derive an asymptotic expression for it.
 }
\maketitle
\section{Introduction}
The importance of symmetries in all branches of physics is undeniable, as both their presence and absence can lead to crucial phenomena~\cite{gross-96}. For instance, when a system with a symmetric action  passes from a symmetric ground state to a less symmetric one, it undergoes spontaneous symmetry breaking. 
This phenomenon has garnered significant attention in both low- and high-energy physics over the past century, as it is responsible for phase transitions and mass generation.
Another intriguing scenario occurs when symmetries that are preserved in classical physics do not always hold once the theory is quantised. 
This discrepancy leads to important and sometimes unexpected consequences, known as quantum anomalies \cite{ach-22}.
On the other hand, the presence of symmetries can also put constraints on a problem. This is what happens in the study of entanglement in the presence of a global conserved charge: the entanglement distributes among the different charge sectors of the theory and it is useful to analyse how the contributions from each sector determine the total entanglement~\cite{laflorencie, gs-18, xavier}. 
This is the essence of the symmetry resolution of  entanglement, which was also motivated by the experimental work~\cite{lukin}, where it is shown that the dynamics of the entanglement in non-equilibrium many-body systems can be better understood by exploiting the internal symmetries. The interplay between entanglement and symmetries has been intensively analysed in recent years in many different systems and contexts~\cite{cgs-18, mdgc-20, crc-20, mdc-21, chen-21, cc-21, mbc-21, eimd-21, acdgm-22, chen-22, ghasemi-23, fmc-23, fac-23, gy-23, dgmnsz-23, northe-23, kmop-23, rach-24, cass-24, bamc-24,mt-22,cmc-23,mdc-24,hc-20,mrc-20,brc-19}. 
Experimental results for symmetry resolved entanglement have also been reported \cite{ncv-21,vek-21,rvm-22}.

The focus of this manuscript is leveraging tools from the theory of entanglement to study symmetry breaking. We consider the entanglement asymmetry, a recently introduced observable that measures how much a symmetry is broken in a portion of an extended quantum system. 
In the original paper~\cite{amc-23}, this quantity is used to investigate the local restoration of a $U(1)$ symmetry after a sudden global quantum quench of a closed system from non-symmetric initial states. What has been found is that very asymmetric initial states can locally restore the symmetry faster than initial states with a lower amount of asymmetry, a phenomenon which has been dubbed \textit{quantum Mpemba effect}. In this analysis, it is crucial to restrict the attention to a subsystem, which relaxes to a (symmetric) stationary state, and not to the total system, whose asymmetry cannot change as it follows a unitary evolution. Since then, the dynamical restoration of symmetries and the quantum Mpemba effect have been examined using the entanglement asymmetry in several different setups, involving free~\cite{makc-24, carc-24, amvc-23} and interacting integrable models~\cite{rkacmb-23, bkccr-24}, random quantum circuits~\cite{lztz-24, tcdl-24}, dissipative systems and mixed states~\cite{cma-24, avm-24}, or higher dimensions~\cite{yac-24}, not only for continuous symmetries but also for discrete ones~\cite{fac-23m, cm-23}. The quantum Mpemba effect has also been experimentally observed in an ion trap simulator by measuring the entanglement asymmetry~\cite{joshi-24}. The notion of entanglement asymmetry has also been extended to investigate the dynamics of confinement~\cite{Khor-23} as well as to study the dynamical restoration of spatial symmetries~\cite{krb-24}. 

Although lots of works have been dedicated to non-equilibrium problems, investigating the properties of the entanglement asymmetry at equilibrium represents an equally interesting field of research. 
For example, in Ref.~\cite{ampc-23}, it is examined in typical random non-symmetric states, finding that a $U(1)$ symmetry emerges in subsystems smaller than half of the total system while, for larger ones, the asymmetry jumps to a non-zero value that grows logarithmically with the subsystem size.
For matrix product states, it has been shown that the
entanglement asymmetry of a generic compact Lie group exhibits a leading-order logarithmic growth with the size of the subsystem and a coefficient fixed by the dimension of the group under study~\cite{cv-23}. A similar leading order behaviour can be also found in the ground state of one dimensional critical systems described by a conformal field theory (CFT), as shown in~\cite{fadc-24}. In this case, the distinctive signature of criticality in the entanglement asymmetry is a correction in the subsystem size $\ell$ of the form $\log\ell/\ell$. This term is semi-universal in the sense that it does not only depends on the CFT data but also on the specific microscopic model.
The entanglement asymmetry can be obtained from the charged moments of the reduced density matrix, which can be identified with the partition functions of the field theory on Riemann surfaces  with multiple non-topological defect lines associated with elements of the group inserted at their branch cuts. These partition functions depend on the specific CFT and symmetry group considered and their calculation has to be worked out case by case. In Ref.~\cite{fadc-24}, the entanglement asymmetry is exactly calculated for the critical XY spin chain, whose continuum limit is described by the massless Majorana fermion, and breaks explicitly the $U(1)$ particle number symmetry. Also Ref.~\cite{chen-23} analyses the entanglement asymmetry for this $U(1)$ symmetry in certain coherent states of the massless compact boson. 
The present manuscript fits into this context, since our goal is to study the entanglement asymmetry in the ground state of a specific model, the critical XXZ spin chain, that corresponds to a massless compact boson in the continuum limit. 
This model has an isotropic point which is invariant under  the $SU(2)$ group of spin rotations.
Away from this point, the symmetry reduces to the $U(1)$ subgroup of rotations around a transverse axis. 
We first study the breaking of the $U(1)$ symmetry of rotations around the longitudinal axis away from the $SU(2)$-symmetric point. 
For this purpose, we exploit the field theory description of the critical XXZ spin chain, in which the $SU(2)$-symmetric point corresponds to the self-dual radius compact boson. 
By performing a perturbative analysis near this point, we calculate the charged moments of the $U(1)$ subgroup of longitudinal rotations. 
This theoretical approach is complemented by a numerical lattice study employing tensor network techniques.
We also examine the entanglement asymmetry of the full $SU(2)$ group of spin rotations. 
By using general scaling arguments and the properties of the charged moments, we derive an asymptotic expression for these moments as a function of the subsystem size. 
This allows us to determine the corresponding entanglement asymmetry, thereby extending the findings of Ref.~\cite{fadc-24}. 
Unlike the symmetry breaking considered in Ref.~\cite{fadc-24}, which involves a continuous group breaking into a finite discrete subgroup, our study focuses on the breaking into a continuous subgroup.

The paper is organised as follows. In Sec.~\ref{sec:defs}, we introduce the definition of the entanglement asymmetry and we express it in terms of the charged moments, which are the main actors of the following parts. Sec.~\ref{sec:model} contains a detailed description of the critical XXZ spin chain and of its continuum counterpart, setting the stage and the main tools we use. Sec.~\ref{sec:U(1)} is devoted to the breaking of the $U(1)$ subgroup of rotations around the longitudinal axis. In Sec.~\ref{sec:charged_moments}, we calculate their charged moments in the underlying CFT close to the symmetric point, which we support with numerical simulations on the lattice. From these results, we compute the corresponding entanglement asymmetry in Sec.~\ref{sec:asymm}. In Sec.~\ref{sec:SU2}, we study the symmetry breaking from the full $SU(2)$ group to the subgroup $U(1)$ of rotations around the transverse axis. We conjecture an expression for the charged moments from which we derive the entanglement asymmetry. We finally draw our conclusions in Section \ref{sec:concl}. We also include two appendices with the details of our analytical and numerical computations.

\section{Entanglement asymmetry as distance between two states}\label{sec:defs}

The idea of the entanglement asymmetry appeared in Ref.~\cite{amc-23} and, as we already stressed in the introduction, its goal is to provide a purely information-theoretical tool to quantify the amount of symmetry breaking in a subsystem of an extended quantum system. Despite its relevance, this issue has not received much attention in the past. The main tools employed for discerning if a symmetry is broken have been local order parameters: these are any operators $O(x)$ that transform non-trivially under the symmetry group $G$ of interest such that, in the presence of the symmetry, one can guarantee that their average is zero, i.e. $\braket{O(x)}=\int_G\mathrm{d} g\braket{U_gO(x)U_g^{\dagger}}=0$, where $\mathrm{d} g$ denotes the Haar measure over the group $G$ and $U_g$ is the unitary operator associated to the group element $g\in G$. Therefore, the order parameter can detect the symmetry breaking when its mean value deviates from zero, or equivalently, when the two-point correlator $\braket{O(0)O(x)}$ has a non-zero asymptotic value for large distances. However, even though a non-vanishing order parameter signals
that the symmetry is broken, the opposite is generally not true.
The entanglement asymmetry bypasses this problem since it univocally determines whether the symmetry is broken or not. 

Let us consider the set of all the states of a system and the convex submanifold of $G$-invariant states, i.e. the states $\tilde{\rho}$ such that $ U_g\tilde{\rho} U_g^{\dagger}=\tilde{\rho}, \,\,\forall g \in G$. The amount the state $\rho$ breaks the symmetry $G$ may be quantified by measuring its distance from this manifold. One may choose many metrics to compute this distance and, in general, it can be very tough to perform the calculation analytically for many-body quantum states. In order to define the entanglement asymmetry, our choice is to use a pseudo-metric borrowed from information theory, the \textit{relative entropy}.
Given two states $\rho$ and $\sigma$, the relative entropy of $\rho$ with respect to $\sigma$ measures the distinguishability between them and it is defined as
\begin{equation}
S(\rho||\sigma)=-\mathrm{Tr}[\rho(\log \sigma-\log\rho) ].
\end{equation}
Contrarily to standard metrics, this quantity is not symmetric, but it satisfies other relevant properties~\cite{nc}; for instance, it is non-negative, $S(\rho||\sigma)\geq 0$, and it decreases under quantum operations.

For a generic compact Lie group $G$, we can write the projection $\rho_G$ of $\rho$ into the space of $G$-invariant states as the outcome of a ``twirling operation'' on $\rho$~\cite{bbpssw-96, bdvsw-96}, i.e. 
\begin{equation}
\rho_G=\frac{1}{{\rm vol}\,G}\int_G\mathrm{d} g U_g\rho U^{\dagger}_g,\label{eq:twirling}
\end{equation}
where ${\rm vol}\,G$ is the volume of the group.
If we take the relative entropy of $\rho$ with respect to $\rho_G$, we define the asymmetry as
\begin{equation}
\Delta S(\rho)= S(\rho||\rho_G).\label{eq:whatisasy}
\end{equation}
This notion of ``asymmetry'' was already introduced from a perspective orthogonal to ours in the context of quantum information and resource theory~\cite{vawj-08,gms-09,t-19,ms-14}. Apart from being non-negative $\Delta S(\rho)\geq 0$ by definition, it vanishes, $\Delta S(\rho)= 0$, if and only if $\rho=\rho_G$. Moreover, $\rho_G$ is the $G$-invariant state with the smallest relative entropy with respect to $\rho$ \cite{gms-09}.

Inserting the explicit expression~\eqref{eq:twirling} of $\rho_G$, using the cyclic property of the trace and recalling the definition of the von Neumann entropy, $S(\rho)=-\Tr(\rho\log\rho)$, we can rewrite Eq.~\eqref{eq:whatisasy} as
\begin{equation}\label{eq:DeltaS}
\Delta S(\rho)=-\mathrm{Tr}[\rho(\log\rho_G-\log\rho)]= S(\rho_G)-S(\rho).
\end{equation}
The appearance of the logarithm of an operator makes the calculation of the asymmetry very challenging. A way to bypass such complication is to introduce the R\'enyi asymmetry,
\begin{equation}
\Delta S_n(\rho)=S_n(\rho_{G})-S_n(\rho), \qquad S_n(\rho)=\frac{1}{1-n}\log \mathrm{Tr}\rho^n,\label{eq:renyiasymm}
\end{equation}
which satisfies similar properties to Eq.~\eqref{eq:DeltaS}, i.e. 
$\Delta S_n(\rho)\geq 0$ and $\Delta S_n(\rho)=0$ only if $\rho=\rho_{G}$, and yields the asymmetry~\eqref{eq:DeltaS} 
in the limit  $n\to 1$~\cite{mhms-22, hms-23}. For integer $n\geq 2$, the R\'enyi 
asymmetry is not only easier to calculate than Eq.~\eqref{eq:DeltaS} but it is also accessible experimentally using for example randomised measurements, as shown recently in an ion trap setup~\cite{joshi-24}.

As we already mentioned, in extended quantum systems, symmetry breaking
is tied to the subsystem of interest. Let us take a system that can be divided into two spatial regions $A$ and $\bar{A}$. We assume that the unitary operator $U_g$ acting on the total Hilbert space $\mathcal{H}=\mathcal{H}_{A}\otimes \mathcal{H}_{\bar{A}}$ can be decomposed as $U_g=U_{g}^{A}\otimes U_{g}^{\bar{A}}$, where $U_{g}^A$ ($U_g^{\bar{A}}$) acts on $\mathcal{H}_A$ ($\mathcal{H}_{\bar{A}}$). 
In this scenario, if the total system is in a state $\rho$, the amount the symmetry $G$ is broken in the subsystem $A$ is given by the R\'enyi \textit{entanglement asymmetry} $\Delta S_A^{(n)}:=\Delta S_{n}(\rho_A)$, where $\rho_A$ is the reduced density matrix  $\rho_A=\Tr_{\bar{A}}(\rho)$, and its twirling $\rho_{A, G}$ is calculated using $U_{g}^{A}$ in Eq.~\eqref{eq:twirling}.

In this manuscript, we will mostly concentrate on the breaking of a $U(1)$ symmetry generated by a charge $Q$ that splits in the charges of $A$ and $\bar{A}$ as $Q=Q_A\otimes \mathbb{I}+\mathbb{I}\otimes Q_{\bar{A}}$. In that case, the action of the elements of this group in the subsystem $A$ is implemented by the unitary operators $U_{\alpha}^A=e^{i\alpha Q_A}$ with $\alpha\in[0, 2\pi)$.
Specialising the expression~\eqref{eq:twirling} of $\rho_{A,G}$ for the $U(1)$ case, we can rewrite the Rényi entanglement asymmetry in terms of the \textit{charged moments} $Z_n(\boldsymbol{\alpha})$,
\begin{equation}
Z_n(\boldsymbol{\alpha})=
 \mathrm{Tr}\left[\prod_{j=1}^n\rho_A e^{i\alpha_{j,j+1}Q_A}\right],\label{eq:asymoments}
\end{equation}
as
\begin{equation}
\Delta S_A^{(n)}=\frac{1}{1-n}\log\left[\displaystyle \int_{-\pi}^{\pi}\frac{\mathrm{d}\alpha_1...\mathrm{d}\alpha_n}{(2\pi)^{n}}\frac{Z_n(\boldsymbol{\alpha})}{Z_n(\boldsymbol{0})}\right],
\end{equation}
with $\alpha_{i,j}:=\alpha_i-\alpha_j$, $\alpha_{n+1}=\alpha_1$ and $\boldsymbol{\alpha}=\{\alpha_1,\dots,\alpha_n\}$.  When $\rho_A$ is a $U(1)$ invariant state, we have that $Z_n(\boldsymbol{\alpha})=Z_n(\boldsymbol{0})$ and, therefore, $\Delta S_A^{(n)}=0$. In field theory,
the charged moment $Z_n(\boldsymbol{\alpha})$ is the partition function of the theory on a $n$-sheet Riemann surface in which the operators $e^{i\alpha_{jj+1}Q_A}$ correspond to (non-topological) defect lines inserted along the branch cuts of the surface~\cite{fadc-24}. Moreover, since $Q$ is local, the charged moments can be calculated efficiently using tensor network techniques.  Therefore, Eq.~\eqref{eq:asymoments} is our starting point to compute the entanglement asymmetry and it will represent the main focus of the following parts of the manuscript.

\section{The XXZ spin chain and its continuum limit}\label{sec:model}

In this section, we introduce the system that we will study in the rest of the manuscript and the CFT that describes its continuum limit. The XXZ spin chain is a one-dimensional spin-$\frac{1}{2}$ chain described in the thermodynamic limit by the Hamiltonian 
\begin{equation}
H=J\sum_{j\in \mathbb Z} \left(S^x_j S^x_{j+1} + S^y_j S^y_{j+1} + \Delta S^z_j S^z_{j+1}\right),\label{eq:xxzham}
\end{equation}
where $S_j^{\mu}=\sigma_j^\mu/2$, $\mu=x, y, z$ and $\sigma_j^\mu$ denote the Pauli operators at the $j$-th site. 
The parameter $J$ is simply an energy scale for the system and, in the following, we take $J>0$. 
The Hamiltonian~\eqref{eq:xxzham} possesses a $U(1)$ rotational symmetry around the $z$ axis, which enlarges to the full $SU(2)$ group of spin rotations when $\Delta=\pm 1$.
In the thermodynamic limit and at zero temperature, the system undergoes two quantum phase transitions when $\Delta=\pm 1$: 
in the critical region $|\Delta|\leq 1$, the Hamiltonian is gapless and, for $-1<\Delta\leq 1$,  in the continuum limit it can be described by a CFT, the Luttinger liquid or equivalently the compact free boson. 
For $|\Delta|>1$, the spectrum is gapped and the model enters in a ferromagnetic ($\Delta<-1$) or an antiferromagnetic ($\Delta>1$) phase. 
Only the transition at $\Delta = 1$, which corresponds to the XXX spin chain, can be described by a conformal field theory, being of the Berezinskii-Kosterlitz-Thouless type~\cite{giamarchi}.

The XXZ Hamiltonian is the paradigmatic example of an interacting integrable model~\cite{gaudin}. 
However, here we do not exploit the integrability of the system but we leverage the low-energy field theory description of the model around the point $\Delta=1$. 
We also mention that a Jordan-Wigner transformation maps the Hamiltonian~\eqref{eq:xxzham} into a fermionic one with a four fermion interaction  for $\Delta\neq 0$.

For this purpose, let us introduce the continuum limit of the XXZ spin chain~\eqref{eq:xxzham} along the critical line $-1<\Delta\leq 1$, which corresponds to the CFT of a free compact boson $\Phi$. 
Its Hamiltonian reads~\cite{affleck-85, lukyanov-98, lt-03}
\begin{equation}\label{eq:xxz_c}
   H= \frac{v}{2}\int \mathrm{d}x \left[K(\partial_x\Theta)^2+\frac{1}{K}(\partial_x\Phi)^2\right],
\end{equation}
where $\partial_x\Theta=-\frac{1}{v}\partial_t\Phi$ plays the role of the conjugate momentum, $v$ is the speed of sound and $K$ is known as Luttinger parameter. These parameters can be expressed in terms of the coupling $\Delta$ in Eq. \eqref{eq:xxzham} as \cite{lukyanov-98, lt-03} 
\begin{equation}
    v=\frac{\pi}{2}\frac{\sqrt{1-\Delta^2}}{\arccos\Delta}, \qquad K=\frac{\pi}{2}\frac{1}{\pi-\arccos\Delta}.
\end{equation}
The dictionary which relates the spin operators in the Hamiltonian \eqref{eq:xxzham} to the fields in Eq. \eqref{eq:xxz_c} is \cite{affleck-85, lukyanov-98, lt-03}
\begin{equation}\label{eq:dictionary}
    \begin{split}
        S_j^z\simeq & \,\frac{a}{\sqrt{\pi}} \partial_x \Phi(x) +c_1\sin(2\sqrt{\pi}\Phi(x))+\cdots,\\
        S_j^x\simeq &\,c_2(-1)^j\cos(\sqrt{\pi}\Theta(x))+\cdots,
    \end{split}
\end{equation}
where the ellipses denote fields with subleading scaling dimension, $a$ is the lattice spacing and $c_{1,2}$ are non-universal coefficients. 
For later convenience, we can introduce the imaginary time $\tau$ and define the complex variable $z=x+i\tau$. 
We use $\phi$ and $\bar{\phi}$ to denote the holomorphic and anti-holomorphic components of $\Phi$, such that $\Phi(z,\bar{z}) = \phi(z)+\bar{\phi}(\bar{z})$ and $\Theta(z, \bar{z})=\phi(z)-\bar{\phi}(\bar{z})$. 

The compact boson is a CFT with central charge $c=1$. Its spectrum of primary fields includes $\partial_{z}\Phi$, with conformal dimensions $(1,0)$, and its anti-holomorphic counterpart $\partial_{\bar{z}}\Phi$, with dimensions $(0,1)$. Another important class of fields in this theory are the vertex operators
\begin{equation}
V_{\beta,\bar{\beta}}(z,\bar{ z})=e^{i\beta \phi(z)+\bar{\beta}\bar{\phi}(\bar{z})},
\end{equation}
with conformal dimensions $1/(8\pi)(\beta^2,\bar{\beta}^2)$. Those vertex operators with $\beta=n/R-2\pi m R$ and $\bar{\beta}=n/R+2\pi m R$, where $m$ and $n$ are integer numbers and $R$ is the compactification radius ($K$ is proportional to $R^{-2}$), are primary fields of this CFT. 
An important result that we will frequently employ later is that the $n$-point correlation function of vertex operators on the complex plane is \cite{yellowbook}
\begin{equation}\label{eq:plane}
\langle\prod_{j=1}^n V_{\beta_j, \bar{\beta}_j}(z_j,\bar{z}_j)\rangle_{\mathbb{C}}=\prod_{j<j'}\left(\frac{z_j-z_{j'}}{a}\right)^{\frac{\beta_j\beta_{j'}}{4\pi}}\left(\frac{\bar{z}_j-\bar{z}_{j'}}{a}\right)^{\frac{\bar{\beta}_j\bar{\beta}_{j'}}{4\pi}}, \qquad \mathrm{if}\,\sum_{j=1}^n\beta_j=0.
\end{equation}
If the neutrality condition on the right is not fulfilled, the correlator vanishes. 

The compact boson theory has a $U(1)$ conserved holomorphic current, $\partial \phi$, which in the XXZ spin chain translates into the symmetry under rotations around the $z$ axis, generated by $S_j^z$. At the self-dual radius $R=1/\sqrt{2 \pi}$ ($K=1/2$), there are two additional conserved holomorphic currents, $V_{\pm \sqrt{8\pi},0}$. These vertex operators have conformal dimensions $(1,0)$ and, together with $\partial\phi$, form a spin-1 triplet of generators of a $SU(2)$ symmetry. This special value of the compactification radius corresponds to the isotropic point $\Delta=1$ of the XXZ spin chain, in which the Hamiltonian~\eqref{eq:xxzham} is fully invariant under the $SU(2)$ group of spin rotations, and the currents $V_{\pm \sqrt{8\pi},0}$ are associated with $S_j^\pm=S_j^x\pm i S_j^y$.

In the following sections, we analyse how the symmetry associated with 
the $U(1)$ subgroup of spin rotations generated by the charge $Q=\sum_j S_j^x$ breaks outside the symmetric point $\Delta=1$. This is the first step to study the $SU(2)$ symmetry-breaking away from the isotropic point. To this end, we study the corresponding entanglement asymmetry for an interval $A=[u,v]$ in the ground state of the critical XXZ spin chain close to $\Delta=1$. As we showed in Eq.~\eqref{eq:asymoments}, the entanglement asymmetry can be determined from the charged moments $Z_n(\boldsymbol{\alpha})$ of the reduced density matrix. We will calculate the latter both analytically in the underlying CFT and numerically using tensor networks techniques.  

In the CFT approach, the first difficulty that we encounter when calculating the charge moments is that, using the dictionary~\eqref{eq:dictionary}, the charge $Q_A=\sum_{j\in A} S_j^x$ takes a complicated non-local form in terms of the fields.
The case $n=1$, $Z_1(\alpha)={\rm Tr}(\rho_A e^{i\alpha Q_A})$, which corresponds to the full counting statistics of $Q_A$, has been analysed in Ref.~\cite{collura-essler-groha}. We can follow the same strategy as in that work to get a simpler expression for the charge.
We perform a rotation of the XXZ Hamiltonian~\eqref{eq:xxzham} choosing the anisotropy axis along the $x$ direction, 
\begin{equation}\label{eq:xxz_rot}
    H=J\sum_j \left(\tilde{S}^x_j \tilde{S}^x_{j+1} + \tilde{S}^y_j \tilde{S}^y_{j+1} +\tilde{S}^z_j \tilde{S}^z_{j+1}+ (\Delta-1) \tilde{S}^x_j \tilde{S}^x_{j+1}\right).
\end{equation}
Its continuum limit is the perturbed compactified boson~\cite{affleck-85, lukyanov-98, lt-03}
\begin{equation}\label{eq:action_rot}
   H= \frac{v}{2}\int \mathrm{d}x \left[(\partial_x\tilde{\Theta})^2+(\partial_x\tilde{\Phi})^2\right]-2\pi v\int \mathrm{d}x \sum_a \,g_a J_a\bar{J}_a,
\end{equation}
where $g_a\propto 1-\Delta$ and $J_a$ ($\bar{J}_a$) are the (anti-)holomorphic $SU(2)$ currents. 
Decomposing the field $\tilde{\Phi}(z, \bar{z})$ into its holomorphic and anti-holomorphic parts, $\tilde{\Phi}(z, \bar{z})=\varphi(z)+\bar{\varphi}(\bar{z})$, we can express the currents $J_a$ and $\bar{J}_a$ in terms of them as
\begin{equation}\label{eq:JJ}
\begin{split}
    J_3(z)=\frac{i}{\sqrt{2\pi}}\partial\varphi(z), \quad \qquad &\bar{J}_3(\bar{z})=\frac{i}{\sqrt{2\pi}}\partial\bar{\varphi}(\bar{z}),\\
J^{\pm}(z)=\frac{1}{2\pi a}e^{\pm i\sqrt{8\pi}\varphi(z)}, \qquad &\bar{J}^{\pm}(\bar{z})=\frac{1}{2\pi a}e^{\mp i\sqrt{8\pi}\bar{\varphi}(\bar{z})}.
   \end{split} 
\end{equation}
The $U(1)$ symmetry of the XXZ Hamiltonian
constrains the coupling constants $g_i$ such that $g_2=g_3$. 

Using the rotation performed in Eq. \eqref{eq:xxz_rot}, the magnetization along the $x$-axis, $S_j^x$, is turned into $\tilde{S}_j^z$, and taking into account that $\tilde{S}_j^z\sim a\partial_x \tilde{\Phi}(x)/\sqrt{2\pi}+\cdots$,  we obtain
\begin{equation}
Q_A=\sum_{j\in A} S_j^x=\sum_{j\in A} \tilde{S}_j^z\simeq \frac{1}{\sqrt{2\pi}}(\tilde{\Phi}(u)-\tilde{\Phi}(v)).
\end{equation}
Therefore, the operators $e^{i\alpha Q_A}$ entering in the charged moments~\eqref{eq:asymoments} correspond in the field theory~\eqref{eq:action_rot} to insert a pair of vertex opertors at the end-points of the interval $A=[u, v]$, 
\begin{equation}\label{eq:chargex}
   e^{i\alpha\sum_{j\in A}S^x_j}%=e^{i\alpha\sum_{j\in A}\tilde{S}^z_j}
   \simeq e^{i\frac{\alpha}{\sqrt{2\pi}}[\tilde{\Phi}(u)-\tilde{\Phi}(v)]}.
\end{equation}
Moreover, through Eq.~\eqref{eq:JJ}, we can write down the total action of the compact boson with $K=1/2$ and subject to the perturbation~\eqref{eq:action_rot} as 
\begin{equation}\label{eq:action_pert}
\begin{split}
    S&=\frac{1}{2}\int {\rm d}^2z (\partial_{\mu}\Phi)^2+S_{\mathrm{int}},\\
    S_{\mathrm{int}}&=\int {\rm d}^2z \left[-g_3\,\partial_z\varphi\partial_{\bar{z}}\bar{\varphi}+
\frac{g_1+g_2}{4\pi a^2}\cos(\sqrt{8\pi}(\varphi+\bar{\varphi}))+\frac{g_1-g_2}{4\pi a^2}\cos(\sqrt{8\pi}(\varphi-\bar{\varphi}))\right].
\end{split}
\end{equation}
As a consequence, if we are interested in the correlation function of a
product of operators $X$, we find at second order in perturbation theory in the couplings $g_i$'s
\begin{equation}\label{eq:exp}
    \braket{X}=\braket{X}_{0}-\braket{X S_{\mathrm{int}}}_0+\frac{1}{2}(\braket{X S_{\mathrm{int}}^2}_0-\braket{X}_0\braket{S_{\mathrm{int}}^2}_0)+...
\end{equation}
where $\braket{\cdot}_0$ denotes an expectation value with respect to the ($SU(2)$-symmetric) unperturbed action.
This last result is the crucial ingredient for the following sections, where we compute the entanglement asymmetry using the expression for the charge operator \eqref{eq:chargex} and the perturbative expansion in Eq. \eqref{eq:exp}.

\section{Symmetry breaking of an Abelian group}\label{sec:U(1)}

In this section, we study the breaking of the $U(1)$ symmetry generated by the charge $Q_A=\sum_{j\in A} S_j^x$ in the ground state of the critical XXZ spin chain~\eqref{eq:xxzham} when $\Delta<1$. We will first compute the corresponding charged moments~\eqref{eq:asymoments} employing the perturbative CFT description~\eqref{eq:action_pert} of the chain around the point $\Delta=1$.  After presenting our analytical results, we validate them through numerical computations and we calculate from them the entanglement asymmetry.

\subsection{Charged moments}\label{sec:charged_moments}

\subsubsection{Perturbative CFT Approach}\label{sec:cft_calc} 
\begin{figure}[t!]
    \centering
    \includegraphics[width=8.5cm]{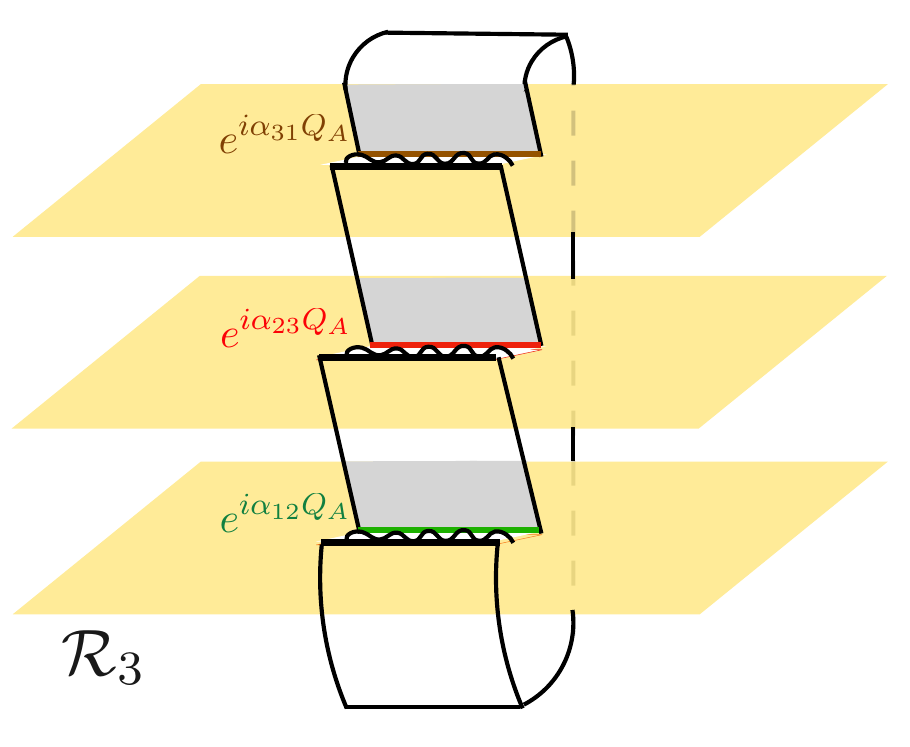}
    \caption{Riemann surface $\mathcal{R}_3$ arising in the 
    calculation of the charged moment $Z_3(\boldsymbol{\alpha})$ in Eq. \eqref{eq:correlator}. The operators $e^{i\alpha_{j,j+1} Q_A}$ are inserted along the interval $A$ in each sheet. For a compact boson, this amounts to inserting two vertex operators at the end-points of the subsystem. }
    \label{fig:riemann}
\end{figure}
Let us consider as subsystem an interval $A=[u, v]$ of length $\ell=|u-v|$ in an infinite line. In this case, using the path integral representation of the reduced density matrix $\rho_A$~\cite{hlw-94, cc-04}, the charged moment $Z_n(\boldsymbol{\alpha})$ is equivalent to the partition function of the field theory~\eqref{eq:action_pert} defined on the $n$-sheet Riemann surface $\mathcal{R}_{n}$ depicted in Fig. \ref{fig:riemann}. The sheets of this surface are sewn cyclically along the branch cut $[u, v]$ as shown in the figure. Taking into account Eq.~\eqref{eq:chargex}, the operator $e^{i \alpha_{j, j+1} Q_A}$ in $Z_n(\boldsymbol{\alpha})$ corresponds to inserting the pair of vertex operators $\tilde{V}_{\beta_j}(u_j)\equiv e^{i\beta_j \tilde{\Phi}(u)}$ and $\tilde{V}_{-\beta_j}(v_j)\equiv e^{-i\beta_j \tilde{\Phi}(v)}$, with $\beta_j=\alpha_{j,j+1}/(\sqrt{2\pi})$, at the end points of the branch cut in the sheet $j=1, \dots, n$. Therefore, within this approach, the charged moment $Z_n(\boldsymbol{\alpha})$ is the $2n$-point correlation function of vertex operators in the field theory~\eqref{eq:action_pert} on the Riemann surface $\mathcal{R}_n$
\begin{equation}
\frac{Z_n^{\rm CFT}(\boldsymbol{\alpha})}{Z_n^{\rm CFT}(\boldsymbol{ 0})}=\langle\prod_{j=1}^n \tilde{V}_{\beta_j}(u_j) \tilde{V}_{-\beta_j}(v_j)\rangle_{\mathcal{ R}_n}.\label{eq:correlator}
\end{equation}
We will calculate this correlator applying the perturbative expansion in Eq.~\eqref{eq:exp} on the Riemann surface $\mathcal{R}_n$, with $X=\prod_{j=1}^n \tilde{V}_{\beta_j}(u_j) \tilde{V}_{-\beta_j}(v_j)$. We start by computing it for the symmetric case in Eq.~\eqref{eq:action_pert}, which corresponds to the $0$-th order term in the expansion~\eqref{eq:exp}, and then we will extend it to the symmetry broken one, by considering the next order terms in~\eqref{eq:exp}.

\textbf{Warm up: Symmetric case}. In the $SU(2)$ symmetric case, i.e. when the couplings $g_i$'s appearing in Eq.~\eqref{eq:action_pert} satisfy $g_i=0$, $i=1,2,3$, we know that $Z_n(\boldsymbol{\alpha})=Z_n(\boldsymbol{0})$ for any $\boldsymbol{\alpha}$.  
To see this from the correlation function~\eqref{eq:correlator} on the Riemann surface $\mathcal{R}_n$, we first map it to the complex plane via the uniformisation map \cite{lm-01, cc-04}
\begin{equation}\label{eq:unif}
z'=\left(\frac{z-u}{z-v}\right)^{1/n}.
\end{equation}
Since the vertex operators in Eq.~\eqref{eq:correlator} are inserted at the branch points of the surface $\mathcal{R}_n$, where there are conical singularities, it is convenient to slightly move them away from those points, inserting them at $u+i\epsilon$ and $v+i\epsilon$. Each point $u+i\epsilon$ and $v+i\epsilon$ is mapped into $n$ points $u_j$ and $v_j$, $j=1,\dots, n$, in the complex plane, with coordinates 
\begin{equation}
\begin{aligned}
u'_j &\approx\left(\frac{\epsilon}{\ell}\right)^{1/n}e^{2\pi i (j-1)/n},\\ 
v'_j &\approx\left(\frac{\ell}{\epsilon}\right)^{1/n}e^{2\pi i (j-1)/n},\label{eq:vu}
\end{aligned}
\end{equation}
at leading order in $\epsilon$. Once we perform this mapping, we have to evaluate 
\begin{equation}
    \frac{Z_n^{\rm CFT}(\boldsymbol{\alpha})}{Z_n^{\rm CFT}(\boldsymbol{ 0})}=\left(\frac{a}{n\epsilon}\right)^{\frac{1}{2\pi}\sum_{j}\beta_j^2}\langle\prod_{j=1}^n \tilde{V}_{\beta_j}(u_j') \tilde{V}_{-\beta_j}(v_j')\rangle_{0, \mathbb{C}},
    \label{eq:maincorrelator}
\end{equation}
where the first factor comes from the Jacobian of the transformation~\eqref{eq:unif}, while the correlator has to be evaluated on the complex plane $\mathbb{ C}$. When $\ell\gg \epsilon$, the vertex operators 
$\tilde{V}_{\beta_j}$ at the points $u_j'$ can be fused into a single vertex operator $\tilde{V}_{\beta_1+\cdots+\beta_n}$ inserted at the point $z'=0$ while those inserted at the points $v_j'$, $\tilde{V}_{-\beta_j}$, fuse into the vertex operator $\tilde{V}_{-\beta_1-\cdots-\beta_n}$ at $z'=v_1'$.  Therefore, for $\ell\gg \epsilon$,
\begin{equation}
\frac{Z_n^{\rm CFT}(\boldsymbol{\alpha})}{Z_n^{\rm CFT}(\boldsymbol{0})}\simeq 
\left(\frac{a}{n\epsilon}\right)^{\frac{1}{2\pi}\sum_j\beta_j^2}
\langle \tilde{V}_{\beta_1+\cdots+\beta_n}(0)\tilde{V}_{-\beta_1-\cdots-\beta_n}(v_1')\rangle_{0,\mathbb{C}}.
\end{equation}
The explicit calculation of the two-point correlator gives 
\begin{equation}
\frac{Z_n^{\rm CFT}(\boldsymbol{\alpha})}{Z_n^{\rm CFT}(\boldsymbol{0})}\simeq 
\left(\frac{a}{n\epsilon}\right)^{\frac{1}{2\pi}\sum_j\beta_j^2}
\left(\frac{\ell}{\epsilon}\right)^{-\frac{1}{2\pi n}(\sum_j\beta_j)^2}.
\end{equation}
Observe that, if in this expression we impose the condition $\sum_j\beta_j=0$, we do not obtain that $Z_n(\boldsymbol{\alpha})/Z_n(\boldsymbol{0})=1$ as we would expect,
but it diverges as $\epsilon\to 0$. This divergence is due to the insertion of the vertex operators $\tilde{V}_{\pm\beta_j}$ at the branch points $u_j$, $v_j$
of the Riemann surface $\mathcal{R}_n$. We can remove it by regularising the vertex operators at the branch points as~\cite{msdz-17, acdgm-22}
\begin{equation}\label{eq:reg_vertex_op}
\tilde{V}_{\beta}^{(*)}(z)=\lim_{\epsilon\to 0}\left(\frac{n\epsilon}{a}\right)^{\frac{\beta^2}{2\pi}}\tilde{V}_{\beta}(z+i\epsilon).
\end{equation}
If we replace the vertex operators in Eq.~\eqref{eq:correlator} by the ones introduced above, we actually find that 
\begin{equation}
\frac{Z_n^{\rm CFT}(\boldsymbol{\alpha})}{Z_n^{\rm CFT}(\boldsymbol{0})}\simeq 
\left(\frac{\ell}{\epsilon}\right)^{-\frac{1}{2\pi n}(\sum_j\beta_j)^2},\label{eq:zeroth}
\end{equation}
which gives the unity when $\sum_j\beta_j=0$. In what follows, we will
adopt the regularisation prescription~\eqref{eq:reg_vertex_op} for the vertex operators inserted at the branch points.

\textbf{Charged moments in the symmetry broken case}. The result that we have found above for the symmetric case ($g_i=0$) corresponds to the $0$-th order term in the perturbative expansion in Eq.~\eqref{eq:exp}. Let us now compute the higher-order terms, which take into account the symmetry breaking when $g_i\neq 0$. 

In the first order term, $I_1=-\langle X S_{\rm int}\rangle_{0, \mathcal{R}_n}$, due to the neutrality condition of the vertex operators, the only term in the action~\eqref{eq:action_pert} that contributes is the one proportional to the $g_3$ coupling. Taking this into account and using the uniformisation transformation~\eqref{eq:unif} to map the Riemann surface $\mathcal{R}_n$ to the complex plane, we obtain 
\begin{equation}\label{eq:I_1}
I_1= g_3\int_\mathbb{C} {\rm d}^2z \langle X \partial_z\varphi(z)\partial_{\bar{z}}\bar{\varphi}(\bar{z})\rangle_{0, \mathbb{C}}.
\end{equation}
Considering separately the holomorphic and the antiholomorphic parts of the correlator above and contracting the derivative field and the $2n$ vertex operators in $X$, we eventually find 
\begin{equation}\label{eq:fointegral}
I_1=-\frac{g_3}{16\pi^2}\left(\frac{\epsilon}{\ell}\right)^{\frac{(\sum_j\beta_j)^2}{ 2\pi n}}\int_{\mathbb{C}}\mathrm{d}^2 z \left|\sum_{j=1}^n\beta_j\left(\frac{1}{z-u'_j}+\frac{1}{v_j'-z}\right)\right|^2.
\end{equation}
This integral is divergent and needs to be regularised: if we exclude from the integration domain a disk of radius $b$ around each of the branch-points of the Riemann surface $\mathcal{R}_n$, then, after the mapping~\eqref{eq:unif}, the integration domain on the complex plane becomes $\sqrt[n]{b/\ell}\leq |z'|\leq\sqrt[n]{\ell/b}$. The integral thus is convergent and, assuming $\epsilon\ll b\ll\ell$, we obtain in Appendix~\ref{app:fo_int} its asymptotic behaviour for large $\ell$, 
\begin{equation}
I_1=-\frac{g_3}{4\pi n}\left(\frac{\epsilon}{\ell}\right)^{\frac{(\sum_j\beta_j)^2}{2\pi n}}\bigg|\sum_{j=1}^n\beta_j\bigg|^2\log\left(\frac{\ell}{b}\right)+\mathcal O(1).\label{eq:fores}
\end{equation}
This expression is zero if $\sum_j \beta_j=0$, which is the condition
that the vertex operators that enter in the calculation of the charged moments must satisfy. Therefore, the leading contribution to the charged moments comes from the second-order perturbation in Eq.~\eqref{eq:exp}. 

The computation of the second-order term, $I_2=(\braket{X S_{\mathrm{int}}^2}_{0,\mathcal{R}_n}-\braket{X}_{0,\mathcal{R}_n}\braket{S_{\mathrm{int}}^2}_{0,\mathcal{R}_n})/2$, is more involved. Taking into account the neutrality condition of the vertex operators, it can be decomposed into three pieces proportional to $g_3^2$, $(g_1+g_2)^2$, and $(g_1-g_2)^2$,
\begin{equation}\label{eq:so_term}
     I_2=\frac{1}{2}\left(g_3^2P_3+\left(\tfrac{g_1+g_2}{4\pi}\right)^2P_++\left(\tfrac{g_1-g_2}{4\pi}\right)^2P_-\right).
\end{equation}
After the uniformisation map~\eqref{eq:unif}, the term $P_3$ reads
\begin{multline}\label{eq:P_3}
P_3=\int_\mathbb{C} {\rm d}^2 z {\rm d}^2w \left[\langle X \partial \varphi(z)\bar{\partial}\bar{\varphi}(\bar{z}) \partial \varphi(w)\bar{\partial}\bar{\varphi}(\bar{w})\rangle_{0,\mathbb{C}}\right. \\ \left.-\langle X\rangle_{0,\mathbb{C}}\langle \partial \varphi(z)\bar{\partial}\bar{\varphi}(\bar{z}) \partial \varphi(w)\bar{\partial}\bar{\varphi}(\bar{w})\rangle_{0,\mathbb{C}}\right].
\end{multline}
Factorising the first correlator in this integral into its holomorphic and anti-holomorphic parts and performing all the contractions between the fields, we obtain
\begin{multline}
\int_\mathbb{C} {\rm d}^2 z {\rm d}^2w \langle X \partial \varphi(z)\bar{\partial}\bar{\varphi}(\bar{z}) \partial \varphi(w)\bar{\partial}\bar{\varphi}(\bar{w})\rangle_{0,\mathbb{C}}=I_1^2/g_3^2+ P_3'\\+
\int_\mathbb{C} {\rm d}^2 z {\rm d}^2w\langle X\rangle_{0,\mathbb{C}}\langle \partial \varphi(z)\bar{\partial}\bar{\varphi}(\bar{z}) \partial \varphi(w)\bar{\partial}\bar{\varphi}(\bar{w})\rangle_{0,\mathbb{C}},
\end{multline}
where $I_1$ is the first order term~\eqref{eq:I_1} and
\begin{multline}
    P_3'=\left(\frac{\epsilon}{\ell}\right)^{\frac{(\sum_j\beta_j)^2}{ 2\pi n}}\frac{1}{32\pi^3}\int_{\mathbb{C}}\frac{\mathrm{d}^2z\mathrm{d}^2w}{|z-w|^4}\mathrm{Re}\left[(z-w)^2\left(\sum_{j=1}^n\beta_j\left(\frac{1}{z-u'_j}+\frac{1}{v_j'-z}\right)\right)\right. \\ \left.\times\left(\sum_{j=1}^n\beta_j\left(\frac{1}{w-u'_j}+\frac{1}{v_j'-w}\right)\right)\right].\label{eq:P3prime}
\end{multline}
Inserting this result in Eq.~\eqref{eq:P_3}, we therefore find that
\begin{equation}
P_3=I_1^2/g_3^2+ P_3'.
\end{equation}
The leading behaviour of $I_1$ at large
$\ell$, once it is regularised, was derived in Eq.~\eqref{eq:fores}. As we already saw, this integral
cancels due to the condition $\sum_{j=1}^n\beta_j=0$ when integrating the charged moments. In Appendix~\ref{app:so_int}, we show that the contribution of $P_3'$ for $n=2$ is of order smaller than $\mathcal{O}(\ell^0)$. 

The term $P_+$ in Eq.~\eqref{eq:so_term} boils down, once we have performed the uniformisation map~\eqref{eq:unif}, to the integral in the complex plane
\begin{multline}
P_+=\frac{1}{a^4}\int_\mathbb{C} {\rm d}^2 z\,{\rm d}^2w
\left[\langle X \cos\left(\sqrt{8\pi}(\varphi(z)+\bar{\varphi}(\bar{z}))\right)
\cos\left(\sqrt{8\pi}(\varphi(w)+\bar{\varphi}(\bar{w})\right)\rangle_{0,\mathbb{C}}\right. \\
\left. -\langle X\rangle_{0,\mathbb{C}}\langle \cos\left(\sqrt{8\pi}(\varphi(z)+\bar{\varphi}(\bar{z}))\right)
\cos\left(\sqrt{8\pi}(\varphi(w)+\bar{\varphi}(\bar{w})\right)\rangle_{0,\mathbb{C}}\right].
\end{multline}
The contraction of the fields in the correlators inside this integral gives
\begin{equation}\label{eq:int_pp}
P_+=\left(\frac{\epsilon}{\ell}\right)^{\frac{(\sum_j\beta_j)^2}{ 2\pi n}}\int_{\mathbb{C}}\mathrm{d}^2z\mathrm{d}^2w\frac{1}{|z-w|^4}\left(\prod_{j=1}^n\bigg|\frac{z-u'_j}{w-u'_j}\frac{w-v'_j}{z-v'_j}\bigg|^{\alpha_{j,j+1}/\pi}-1\right).
\end{equation}
On the other hand, the term $P_-$ in Eq.~\eqref{eq:so_term} corresponds to
\begin{multline}
P_-=\frac{1}{a^4}\int_\mathbb{C} {\rm d}^2 z\,{\rm d}^2w
\left[\langle X \cos\left(\sqrt{8\pi}(\varphi(z)-\bar{\varphi}(\bar{z}))\right)
\cos\left(\sqrt{8\pi}(\varphi(w)-\bar{\varphi}(\bar{w})\right)\rangle_{0,\mathbb{C}}\right. \\
\left. -\langle X\rangle_{0,\mathbb{C}}\langle \cos\left(\sqrt{8\pi}(\varphi(z)-\bar{\varphi}(\bar{z}))\right)
\cos\left(\sqrt{8\pi}(\varphi(w)-\bar{\varphi}(\bar{w})\right)\rangle_{0,\mathbb{C}}\right],
\end{multline}
which leads to the integral
\begin{equation}\label{eq:int_pm}
P_-=\left(\frac{\epsilon}{\ell}\right)^{\frac{(\sum_j\beta_j)^2}{ 2\pi n}}\int_{\mathbb{C}}\frac{\mathrm{d}^2z\mathrm{d}^2w}{|z-w|^4}\left[\prod_{j=1}^n\left(\frac{z-u'_j}{w-u'_j}\frac{w-v'_j}{z-v'_j}\frac{\bar{w}-u^{'*}_j}{\bar{z}-u^{'*}_k}\frac{\bar{z}-v^{'*}_j}{\bar{w}-v^{'*}_j}\right)^{\frac{\alpha_{jj+1}}{2\pi}}-1\right].
\end{equation}
For a generic integer $n$, it is hard to solve the (divergent) integrals~\eqref{eq:int_pp} and~\eqref{eq:int_pm} explicitly. However, as shown in detail in Appendix~\ref{app:so_int}, we can regularise them for $n=2$, where $\alpha_1=-\alpha_2\equiv \alpha$, and obtain their leading order behaviour in the subsystem size $\ell$,
\begin{equation}\label{eq:Ppm_final}
P_+=-P_-=\frac{\alpha^2}{4}\frac{\epsilon}{b}\log\left(\frac{\ell}{b}\right)+\mathcal O(1).
\end{equation}
This expression is very peculiar because it contains a ratio of the $\epsilon$ and $b$ cutoffs in the coefficient of the leading order term. As a consequence, we cannot predict its exact value from this perturbative approach. 

Summing all the contributions that we have computed and taking into account that $g_j\propto 1-\Delta$, we conclude that, close to the symmetric point $\Delta=1$, the $n=2$ charged moment of the underlying CFT behave as
\begin{equation}\label{eq:resultCFTwow}
\frac{Z_2^{\rm CFT}(\boldsymbol{\alpha})}{Z_2^{\rm CFT}(\boldsymbol{0})}-1\propto (1-\Delta)^2\alpha^2\log\left(\frac{\ell}{b}\right)+\mathcal O((1-\Delta)^3).
\end{equation}
The uncharged moment $Z_2^{\rm CFT}(\boldsymbol{0})$ is the purity of the subsystem $A$ which, for a critical system, scales as $Z_2(\boldsymbol{0})\sim\ell^{-c/4}$ and it only depends on the central charge of the CFT~\cite{hlw-94, cc-04}, which is $c=1$ in our case.
In the next subsection, we numerically investigate the $n=2$ charged moments in the ground state of the critical XXZ spin chain and compare with the result in Eq.~\eqref{eq:resultCFTwow}. 

\subsubsection{Numerical Approach}

The goal of this subsection is to check the prediction that we have found for the $n=2$ charged moments in Eq.~\eqref{eq:resultCFTwow}. In order to do that, we have performed numerical simulations based on an infinite Matrix Product State (iMPS) representation of the infinite spin chain in Eq.~\eqref{eq:xxzham}. The search of its ground state has been carried out with the iDMRG method implemented in the TeNPy library~\cite{tenpy}. Further details can be found in Appendix~\ref{app:numerics}.
Once a good approximation of the ground state is obtained through the iMPS technique, we can compute the charged moments $Z_n(\boldsymbol{\alpha})$ of a subsystem of length $\ell$. 
Since $Q_A=\sum_{j\in A}S^x_j$, the operators $e^{i\alpha_{j,j+1} Q_A}$ that enter in their definition~\eqref{eq:asymoments} are factorised along the site indices, and they can be represented as a matrix product operator of bond dimension one. As we show in Appendix~\ref{app:numerics}, in our algorithm, the main limitation is given by the size of the bond dimension, $\chi$, of the ground state MPS representation we have access to.

We will restrict most of our discussion to the $n=2$ charged moments, $Z_2(\alpha)\equiv Z_2(\alpha,-\alpha)$. 
Before analysing $Z_2(\alpha)$, we briefly review the case $n=1$, which corresponds to the generating function or full counting statistics (FCS) of the charge $Q_A$ and has been extensively studied in Ref.~\cite{collura-essler-groha}. 
Through a combination of analytical and numerical calculations, it has been shown in Ref.~\cite{collura-essler-groha} that $Z_1(\alpha)$ is accurately described by the function 
\begin{equation}
Z_1(\alpha)\approx A_1(\alpha,\Delta,(-1)^\ell)\frac{e^{-\ell/\xi_1(\alpha,\Delta)}}{\ell^{\gamma_1(\alpha,\Delta,(-1)^\ell)}},\label{eq:ansatz-numerics}
\end{equation}
where $(-1)^{\ell}$ takes into account the oscillating behaviour for even and odd values of $\ell$ of the coefficients in this expression. 
This feature is reminiscent of the oscillations found for the generating function of the transverse magnetisation in the ground state of the critical XX spin chain~\cite{collura-essler-groha} and of the non-critical XXZ chain~\cite{ccdgm-20}, as well as the parity effects observed in R\'enyi entropies \cite{ccen-09,ce-10}.
The coefficient $\xi_1(\alpha,\Delta)$ can be interpreted as an effective correlation length of $Z_1(\alpha)$. 
Close to $\Delta=1$, $\alpha=0$, the coefficients $\xi_1(\alpha, \Delta)$ and $\gamma_1(\alpha, \Delta)$ behave as
\begin{equation}
\xi_1^{-1}(\alpha,\Delta)\sim \alpha^2(1-\Delta)^2, \qquad \gamma_1(\alpha,\Delta)\sim \alpha^2(1-\Delta).
\end{equation}
As a consequence, when $\Delta\to 1$, the power-law term in Eq.~\eqref{eq:ansatz-numerics} dominates over the asymptotic exponential decay for a wide range of the subsystem size $\ell$.

\begin{figure}[t]
\includegraphics[width=0.49\textwidth]{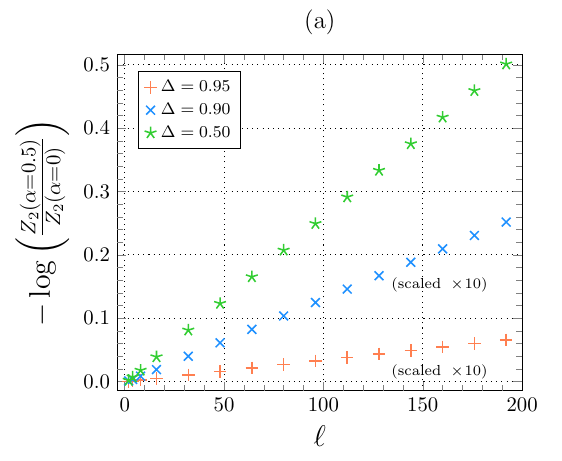}
\includegraphics[width=0.49\textwidth]{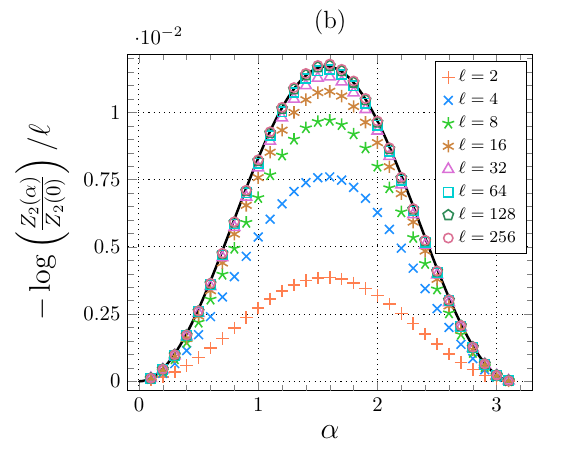}
\caption{Panel (a): We report the numerical values of $-\log\frac{Z_2(\alpha)}{Z_2(0)}$ as a function of the subsystem length $\ell$ in the ground state of the critical XXZ spin chain for $\alpha=0.5$ and different couplings $\Delta$. The points corresponding to $\Delta=0.90$ and $0.95$ are rescaled by a factor $10$. Panel (b): We study $-\log\left(\frac{Z_2(\alpha)}{Z_2(0)})\right)/\ell$ as a function of $\alpha$ for $\Delta=0.5$ and different subsystem lengths $\ell$. As $\ell$ increases, the numerical data tends to the correlation length $\xi_2(\alpha,\Delta)$, see Eq.~\eqref{eq:yayfit}. The solid curve is the function $\xi_2(\alpha, \Delta)=c_2 \sin^2(\alpha)$, with $c_2$ fitted to the points for $\ell=256$.}
\label{fig:linear_charged_mom}
\end{figure}

Let us now move on to the $n=2$ charged moments $Z_2(\alpha)$. 
As we will see, the ansatz~\eqref{eq:ansatz-numerics} also describes well this case when $\ell$ is large. 
In Fig.~\ref{fig:linear_charged_mom}~(a), we study $-\log\frac{Z_2(\alpha)}{Z_2(0)}$ as a function of the subsystem size $\ell$ for $\alpha=0.5$ and different values of the coupling, $\Delta=0.5, 0.9, 0.95$, taking as bond dimension $\chi=100$. All the data points correspond to even subsystem lengths but even/odd effects similar to those observed for $Z_1(\alpha)$ are also present, although we will not study them here. 
From the figure, we can clearly see that $-\log\frac{Z_2(\alpha)}{Z_2(0)}$ grows linearly with $\ell$, indicating that the main deviation of $Z_2(\alpha)$ from $Z_2(0)$ is an exponential decaying function, $e^{-\ell/\xi_2(\alpha,\Delta)}$. The correlation length $\xi_2(\alpha,\Delta)$ decreases as a function of $1-\Delta$, and must diverge for $\Delta =1$, where $Q_A$ is conserved and the charged moments reduce to $Z_2(\alpha)=Z_2(0)$.

To further verify this linear growth, in Fig.\ref{fig:linear_charged_mom}(b), we examine the dependence of the charged moments on $\alpha$ by dividing $-\log \frac{Z_2(\alpha)}{Z_2(0)}$ by the subsystem length $\ell$ for $\Delta=0.5$. 
We take bond dimension $\chi=80$, since a larger value of it does not affect the leading behaviour. The different symbols correspond to increasing subsystem lengths $\ell$. We observe that, for a given $\alpha$, the points tend to converge as $\ell$ increases to a value
that should correspond to $\xi_2^{-1}(\alpha, \Delta)$,  as expected if $-\log\frac{Z_2(\alpha)}{Z_2(0)}\sim \ell/\xi_2(\alpha, \Delta)$.
The solid line is the result of fitting the function $\xi_2^{-1}(\alpha,\Delta)=c_2 \sin^2(\alpha)$ to the points for $\ell=256$. We find that this fit only improves as $-\log \frac{Z_2(\alpha)}{Z_2(0)}$ gets smaller, either by approaching the critical value $\Delta=1$ or by reducing the value of $\ell$. This would mean that the exact $\alpha$-dependence of $\xi_2(\alpha, \Delta)$ has some minor corrections with respect to the ansatz $\xi_2^{-1}(\alpha,\Delta)\propto \sin^2(\alpha)$.

\begin{figure}[t]
\centering 
\includegraphics[width=0.5\textwidth]{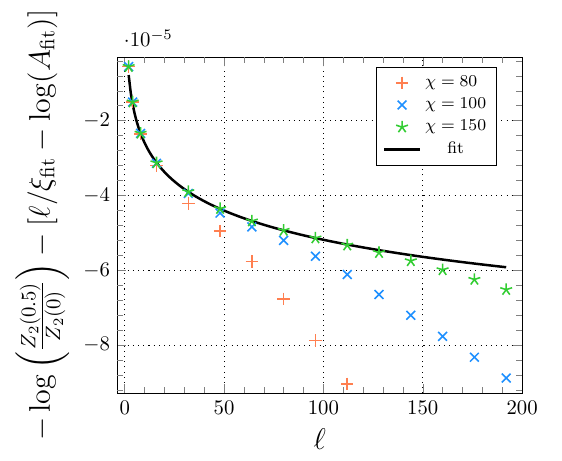}
\caption{The subleading corrections in the subsystem size $\ell$ for $-\log\frac{Z_2(\alpha)}{Z_2(0)}$ with $\alpha=0.5$ and $\Delta=0.95$ are analyzed. Each symbol represents the numerical value of $-\log\frac{Z_2(\alpha)}{Z_2(0)}$ calculated using a different bond dimension $\chi$, after subtracting the linear and $\mathcal O(1)$ terms of the function~\eqref{eq:yayfit}, which was fitted to the numerical results of $-\log\frac{Z_2(\alpha)}{Z_2(0)}$ for $\chi=150$ and subsystem lengths $\ell \in {8, 16, \dots, 96}$. The solid line represents the function $\gamma_2(\alpha, \Delta)\log\ell$, with the coefficient $\gamma_2(\alpha, \Delta)$ obtained from the above fit. }
\label{fig:plot5}
\end{figure}

\begin{figure}[t]
\centering
\includegraphics[width=0.49\textwidth]{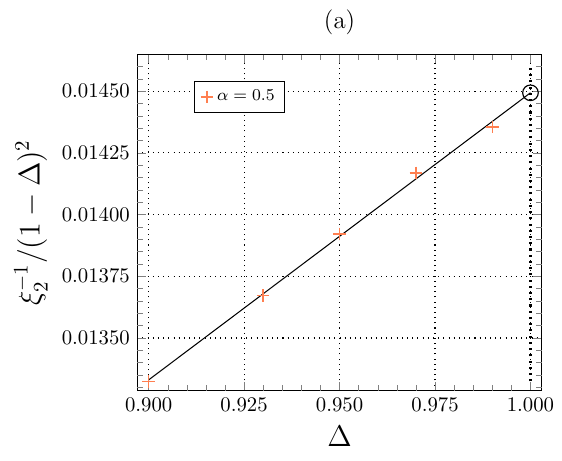}
\includegraphics[width=0.49\textwidth]{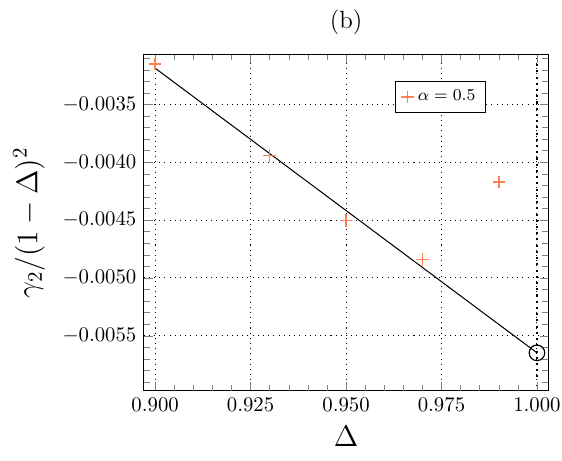}
\includegraphics[width=0.49\textwidth]{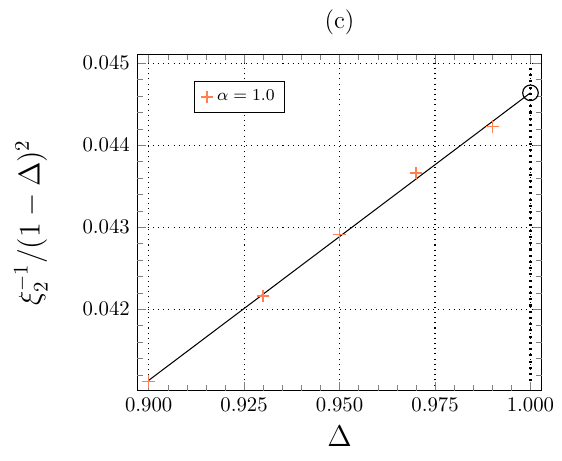}
\includegraphics[width=0.49\textwidth]{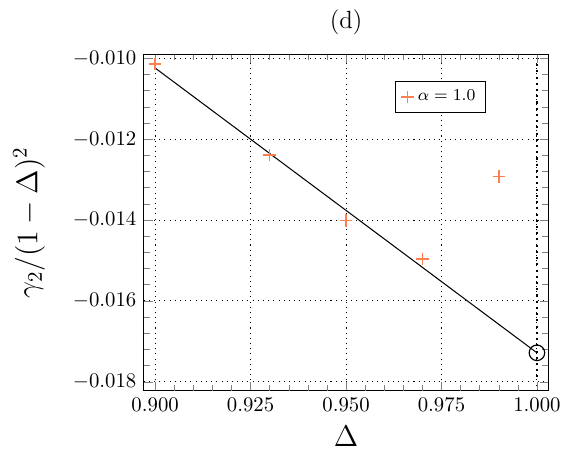}

\caption{The symbols represent the coefficients $\xi_2^{-1}(\alpha,\Delta)$ (left panels) and $\gamma_2(\alpha,\Delta)$ (right panels) obtained in the fit of the function~\eqref{eq:yayfit} to the numerical results of $-\log\frac{Z_2(\alpha)}{Z_2(0)}$ for the set of subsystem lengths $\ell\in\{8, 16, \dots, 96\}$, taking  different values of $\Delta$ and $\alpha=0.5$ (upper panels) or $1$ (lower panels) and bond dimension $\chi=150$. The solid lines represent the function~\eqref{eq:polyfit} fitted  to the numerical data of the corresponding panel, excluding in all cases the point $\Delta=0.99$. }
\label{fig:fits}
\end{figure}

Let us now analyse the subleading corrections of $-\log\frac{Z_2(\alpha)}{Z_2(0)}$ in $\ell$ and their dependence on the coupling $\Delta$. These two aspects are more subtle than those discussed so far, and require more precision to be determined accurately. Therefore, for this analysis we have used a bond dimension $\chi=150$. We conjecture that the quotient $\frac{Z_2(\alpha)}{Z_2(0)}$ has the same asymptotic behaviour as in the case $n=1$, Eq.~\eqref{eq:ansatz-numerics}, and our fitting function is
\begin{equation}\label{eq:yayfit}
-\log\left(\frac{Z_2(\alpha)}{Z_2(0)}\right)\approx \ell/\xi_2(\alpha,\Delta) +\gamma_2(\alpha,\Delta)\log\ell -\log(A_2(\alpha,\Delta)).
\end{equation}
We first take $\Delta=0.95$ and $\alpha=0.5$, with bond dimension $\chi=150$, and we perform a fit of the function~\eqref{eq:yayfit} to the numerical values of $-\log\frac{Z_2(\alpha)}{Z_2(0)}$ obtained for the subsystem lengths $\ell\in\{8,16,..., 96\}$. We plot the results in Fig.~\ref{fig:plot5}. To reveal the subleading behaviour in $\ell$, we have subtracted from all the data points the linear and the constant terms obtained in the fit. The solid line is the function $\gamma_2(\alpha,\Delta)\log\ell$ 
using as coefficient $\gamma_2(\alpha,\Delta)$ the one we found in the fit. To show that the data converges as $\chi$ is increased, we also represent the same points, but calculated for smaller bond dimensions $\chi$, to which we have subtracted the linear and constant terms obtained in the fit for the case $\chi=150$. The numerical data converge as $\chi$ is increased only up to $\ell\lesssim 100$, since larger values of $\ell$ would require larger $\chi$. For other values of $\Delta$, we find that the convergence improves as one gets farther from $\Delta=1$, showing that our numerical simulations become increasingly more difficult as we approach the symmetric point. We have repeated the same fit of the function~\eqref{eq:yayfit} for various couplings $\Delta$, taking $\alpha=0.5$ and $1.0$ and $\ell\in\{8,16,..., 96\}$. The coefficients $\xi^{-1}_2(\alpha,\Delta)$ and $\gamma_2(\alpha,\Delta)$ obtained in the fits are represented in Fig.~\ref{fig:fits}. The results shown in those plots are well described by a simple polynomial expansion of the form
\begin{equation}\label{eq:polyfit}
C_1(\alpha)\cdot(1-\Delta)^2+C_2(\alpha)\cdot(1-\Delta)^3+\mathcal O((1-\Delta)^4).
\end{equation}
Therefore, both $\xi^{-1}_2(\alpha,\Delta)$ and $\gamma_2(\alpha,\Delta)$ are of the same order in $(1-\Delta)$: this explains why, in contrast to the FCS ($n=1$) in Eq.~\eqref{eq:ansatz-numerics}, there is no a range of subsystem lengths $\ell$ in which the logarithmic term in Eq.~\eqref{eq:yayfit} can dominate. We have fitted the function~\eqref{eq:polyfit} to the points of each panel in Fig.~\ref{fig:fits}, excluding in all the cases the point $\Delta=0.99$.
It is reasonable that such a point deviates from the rest since this is the value of $\Delta$ for which the convergence in the bond dimension $\chi$ is worse. In Table~\ref{tab:numfitres}, we report the coefficients  $C_1(\alpha)$ and $C_2(\alpha)$ of the fitting function~\eqref{eq:polyfit} that we have obtained for the data in each panel of Fig.~\ref{fig:fits}. We observe that, in all the rows, the ratio of the coefficients for the two values of $\alpha$ considered is approximately $\frac{\sin^2(1.0)}{\sin^2(0.5)}\approx 3.08$. This would imply that, at leading order, $\gamma_2(\alpha,\Delta)\propto \sin^2(\alpha)$, as in the case of the correlation length $\xi_2^{-1}(\alpha, \Delta)$, see Fig.~\ref{fig:linear_charged_mom}~(b).

\begin{table}[t]
\centering
        \begin{tabular}{rccc|c}\cline{2-5}
        & $\alpha$ & $0.5$ & $1.0$ & ratio \\ \cline{2-5}
        \multirow{2}{*}{$\xi^{-1}\ \bigg\{$} & \multicolumn{1}{r}{$C_1$} & \multicolumn{1}{r}{$ 1.45\cdot 10^{-2}$} & \multicolumn{1}{r|}{$ 4.46\cdot 10^{-2}$} & \multicolumn{1}{r}{$3.08$} \\
                                             & \multicolumn{1}{r}{$C_2$} & \multicolumn{1}{r}{$ 1.16\cdot 10^{-2}$} & \multicolumn{1}{r|}{$ 3.51\cdot 10^{-2}$} & \multicolumn{1}{r}{$3.01$} \\
        \multirow{2}{*}{$\gamma\ \bigg\{$}   & \multicolumn{1}{r}{$C_1$} & \multicolumn{1}{r}{$-5.65\cdot 10^{-3}$} & \multicolumn{1}{r|}{$-1.73\cdot 10^{-2}$} & \multicolumn{1}{r}{$3.06$} \\
                                             & \multicolumn{1}{r}{$C_2$} & \multicolumn{1}{r}{$-2.46\cdot 10^{-2}$} & \multicolumn{1}{r|}{$-7.05\cdot 10^{-2}$} & \multicolumn{1}{r}{$2.86$}\\\cline{2-5}
        \end{tabular}
    \caption{Coefficients of the function~\eqref{eq:polyfit} obtained in the fit to the points of each panel of Fig.~\ref{fig:fits}, excluding in all the cases the point corresponding to $\Delta=0.99$.}
    \label{tab:numfitres}
\end{table}

\begin{figure}[t]
\centering 
\includegraphics[width=0.5\textwidth]{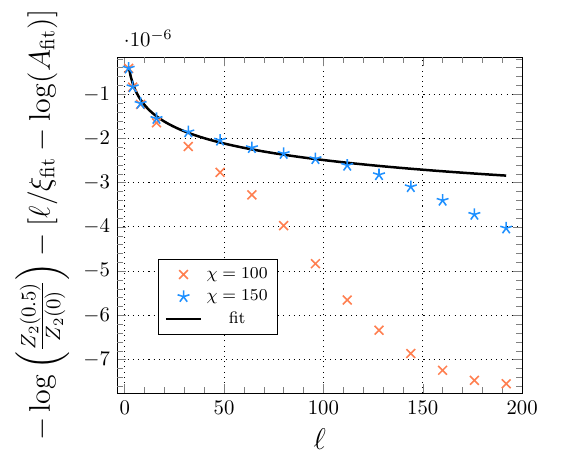}
\caption{Analysis of the subleading corrections in the subsystem length $\ell$ of $-\log\frac{Z_2(\alpha)}{Z_2(0)}$ for $\alpha=0.5$ and $\Delta=0.99$. The symbols correspond to the numerical results obtained for $-\log\frac{Z_2(\alpha)}{Z_2(0)}$ taking two different bond dimensions $\chi$, from which we have subtracted the linear and the $O(1)$ terms of Eq.~\eqref{eq:yayfit} using as coefficients the ones given by the function~\eqref{eq:polyfit} and Table~\ref{tab:numfitres}. The solid line is the function $\gamma_2(\alpha, \Delta)\log\ell$ where $\gamma_2(\alpha, \Delta)$ is the value predicted by Eq.~\eqref{eq:polyfit} and Table \ref{tab:numfitres}}
\label{fig:plot8}
\end{figure}

As a further test, in Fig.~\ref{fig:plot8} we plot the numerical values that we obtain for $-\log\frac{Z_2(\alpha)}{Z_2(0)}$ at  $\Delta=0.99$ and  $\alpha=0.5$ as a function of the subsystem length $\ell$ for two bond dimensions, subtracting the linear and independent terms of the function~\eqref{eq:yayfit}  as in Fig.~\ref{fig:plot5}, but employing for the coefficients $\xi_2(\alpha, \Delta)$ and $\gamma_2(\alpha, \Delta)$ the prediction in Eq.~\eqref{eq:polyfit} with the coefficients in Table \ref{tab:numfitres}. The solid line corresponds to the curve $\gamma_2(\alpha, \Delta)\log \ell$ taking as coefficient $\gamma_2(\alpha, \Delta)$ the value given by Eq.~\eqref{eq:polyfit} and Table~\ref{tab:numfitres}. The agreement of the curve with the  numerical data for bond dimension $\chi=150$ is still good.

\subsubsection{Discussion of the results}\label{sec:discussion}

From the previous numerical analysis, we can conclude that the $n=2$ 
charged moments in the ground state of the critical XXZ spin chain 
are well described by the ansatz~\eqref{eq:yayfit}. However, this 
seems in contradiction with the expression obtained for them in 
Eq.~\eqref{eq:resultCFTwow} using conformal perturbation theory close 
to the symmetric point $\Delta=1$. Let us understand how to reconcile the two results.

According to the discussion at the beginning of Sec.~\ref{sec:cft_calc}, the charged moments $Z_n(\boldsymbol{\alpha})$ can be interpreted as the partition function of the theory on the Riemann surface $\mathcal{R}_n$ with $n$ defect lines inserted along each branch cut. 
If  we impose, far away from the branch cuts, periodic boundary conditions 
in both directions of each sheet of $\mathcal{R}_n$  in order to 
regularise its area $|\mathcal{R}_n|$, then, by general scaling 
arguments, the free energy $-\log Z_n(\boldsymbol{\alpha})$ in a 
critical 1+1 theory should behave as
\begin{equation}\label{eq:free_energy_riem}
-\log Z_n(\boldsymbol{\alpha})\sim f_{{\rm bulk}}|\mathcal{R}_n|+\sum_{j=1}^n t(\alpha_{jj+1}) \ell+\log Z_n^{{\rm CFT}}(\boldsymbol{\alpha}).
\end{equation}
In this expression, $f_{{\rm bulk}}$ is the bulk free energy density and $t(\alpha_{jj+1})\ell$ is the contribution of the defect line corresponding to $e^{i\alpha_{jj+1}Q_A}$, which is proportional to the volume of the defect, in this case $\ell$, and to its tension $t(\alpha_{jj+1})$. These two terms are cut-off dependent, i.e. they depend on the specific microscopic model, and they are not captured by the CFT that describes the low-energy physics of the spin chain. The extra contribution $\log Z_n^{\rm CFT}(\boldsymbol{\alpha})$ arises from the conical singularities at the branch points of $\mathcal{R}_n$. This term is cut-off independent and should be the one compared with the result obtained in Eq.~\eqref{eq:resultCFTwow}. Indeed, it is generically expected to be~\cite{knizhnik-87, dixon-87, cp-88}
\begin{equation}\label{eq:gen_cft_charged_mom}
Z_n^{{\rm CFT}}(\boldsymbol{\alpha})\sim \left(\frac{\ell}{b}\right)^{-\frac{c}{12}(n-1/n)-\gamma_n(\boldsymbol{\alpha}, \Delta)}.
\end{equation}
In our case, the universal exponent $\gamma_n(\boldsymbol{\alpha}, \Delta)$ depends on 
the compactification radius or the Luttinger liquid parameter and, consequently, on the coupling $\Delta$. It vanishes when the defects are topological, i.e. when the corresponding symmetry is respected. 
Therefore, for $\gamma_n(\boldsymbol{\alpha}, \Delta)\ll 1$ close to the point $\Delta=1$, Eq.~\eqref{eq:gen_cft_charged_mom} can be expanded as
\begin{equation}
\frac{Z_n^{\rm CFT}(\boldsymbol{\alpha})}{Z_n^{{\rm CFT}}(\boldsymbol{0})}\sim 1-\gamma_n(\boldsymbol{\alpha}, \Delta)\log\left(\frac{\ell}{b}\right).
\end{equation}
The result found employing perturbative CFT in 
Eq.~\eqref{eq:resultCFTwow} matches this form and we can identify 
$\gamma_2(\alpha, \Delta)\propto (1-\Delta)^2\alpha^2$. 
Indeed, we show in panels (b) and (d) of Fig.~\ref{fig:fits} that, for 
small values of $\alpha$, the universal exponent $\gamma_2(\alpha, \Delta)$ is proportional to $(1-\Delta)^2\alpha^2$ at leading order in 
$1-\Delta$. In addition to this, the numerics in those figures indicate the presence of a rather large third-order correction of the 
form $(1-\Delta)^3\alpha^2$, which should in principle be accessible (but very cumbersome)
by applying the perturbative CFT methods in Sec.~\ref{sec:cft_calc} to 
the next-to-leading-order level. 

\begin{figure}[t]
\centering
\includegraphics[width=0.5\textwidth]{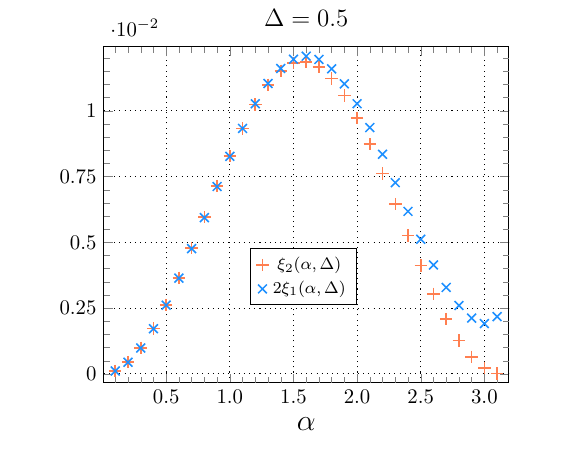}
\caption{Comparison between the inverse correlation lengths $\xi_n^{-1}(\alpha, \Delta)$ of $Z_1(\alpha)$ and $Z_2(\alpha)$ at $\Delta=0.5$. The symbols have been obtained from the 
fit of the functions~\eqref{eq:ansatz-numerics} and~\eqref{eq:yayfit} to the numerical values of $-\log\frac{Z_n(\boldsymbol{\alpha})}{Z_n(\boldsymbol{0})}$ for the set of subsystem lengths $\ell\in\{2, 4, 8, \dots, 256\}$ and bond dimension $\chi=80$. In order to check the identity $\xi_2^{-1}(\alpha,\Delta)=2\xi_1^{-1}(\alpha, \Delta)$ derived in Sec.~\ref{sec:discussion}, the symbols corresponding to $\xi_1^{-1}(\alpha,\Delta)$ have been rescaled by $2$.}
\label{fig:conjecture}
\end{figure}

If we calculate the ratio $Z_2(\alpha)/Z_2(0)$ applying Eqs.~\eqref{eq:free_energy_riem} and~\eqref{eq:gen_cft_charged_mom}, then the term $f_{{\rm bulk}}|\mathcal{R}_n|$ cancels out and we obtain the fitting function~\eqref{eq:yayfit} upon identifying the inverse correlation length $\xi_2^{-1}(\alpha, \Delta)$ with the sum of the tensions $t(\alpha)$, $t(-\alpha)$ of the two defect lines, $\xi_2^{-1}(\alpha, \Delta)=t(\alpha)+t(-\alpha)$. Since $Z_2(\alpha)$ is an even function in 
$\alpha$, this implies that $\xi_2^{-1}(\alpha, \Delta)=2\xi_1^{-1}(\alpha, \Delta)$, where $\xi_1^{-1}(\alpha, \Delta)$ is the inverse correlation length of the FCS $Z_1(\alpha)$, which is the tension of a single defect line. We check this equality in Fig.~\ref{fig:conjecture}. where we plot both  $2\xi_1^{-1}(\alpha,\Delta)$ and  $\xi_2^{-1}(\alpha,\Delta)$ as a function of $\alpha$ for $\Delta=0.5$. The symbols have been obtained by fitting the moments $Z_1(\alpha)$ and $Z_2(\alpha)/Z_2(0)$ to the ansatz in Eqs.~\eqref{eq:ansatz-numerics} and~\eqref{eq:yayfit}, respectively. The overlap is remarkable: the two quantities are almost identical far from $\alpha=\pi$. 
This small discrepancy is due to the fact that $Z_1(\alpha)$ shows a power-law decay at $\alpha=\pi$ for even subsystems sizes~\cite{collura-essler-groha} while $Z_2(\pi)=Z_2(0)$ due to the $\mathbb{Z}_2$ spin flip symmetry. 
Thus, both $\xi_1^{-1}(\pi,\Delta)$ and $\xi_2^{-1}(\pi, \Delta)$ are expected to vanish, but the former has larger correction as seen in the figure. 
This would also explain why in the plots $2\xi_1^{-1}(\alpha, \Delta)$ and $\xi_2^{-1}(\alpha,\Delta)$ become father apart as we approach that point.
We have numerically checked for $\Delta=0.95$ that the $n=3$ charged moments also decay exponentially in $\ell$ with a correlation length satisfying the rule
\begin{equation}
\xi_n^{-1}(\boldsymbol{\alpha},\Delta)=\sum_{j=1}^n\xi_1^{-1}(\alpha_{jj+1},\Delta).
\end{equation}
For the XXZ spin chain, we cannot rigorously prove this equality, since we lack an analytic expression for $\xi_1^{-1}(\alpha,\Delta)$, but it has recently been demonstrated in the ground state of the XY spin chain for the $U(1)$ group of rotations around the $z$ axis~\cite{makc-24}.

\subsection{R\'enyi entanglement asymmetry}\label{sec:asymm} 

From the results of the previous subsection on the $n=2$ charged moments, we can obtain an asymptotic expression for the corresponding 
R\'enyi entanglement asymmetry applying Eq.~\eqref{eq:asymoments}.
We have concluded that the $n=2$ charged moments behave for large subsystem sizes $\ell\gg 1$ as
\begin{equation}
\frac{Z_2(\alpha)}{Z_2(0)}= A_2(\alpha,\Delta) e^{-\ell/\xi_2(\alpha,\Delta)}\ell^{-\gamma_2(\alpha,\Delta)},\label{eq:resultfinalfinal}
\end{equation}
where
\begin{equation}
\xi_2^{-1}(\alpha,\Delta)=\bar{\xi}^{-1}_2(\Delta)\sin^2(\alpha),\qquad \gamma_2(\alpha,\Delta)=\bar{\gamma}_2(\Delta)\sin^2(\alpha).
\end{equation}
Then, according to Eq.~\eqref{eq:asymoments}, we find
\begin{equation}\label{eq:int_charg_mom}
\frac{\mathrm{Tr}(\rho_{A,Q}^2)}{\mathrm{Tr}(\rho_{A}^2)}= \int_{-\pi}^{\pi}\frac{\mathrm{d}\alpha}{2\pi}A_2(\alpha,\Delta) e^{-(\ell/\bar{\xi}_2(\Delta)+\bar{\gamma}_2(\Delta)\log\ell)\sin^2(\alpha)}.
\end{equation}
By taking the large-$\ell$ limit, we can solve the integral through the saddle-point approximation. The saddle points are $\alpha_0=0,\pi$, which correspond to the elements of the subgroup $\mathbb{Z}_2$ of spin flips around the $x$ axis that remain a symmetry of the XXZ spin chain outside the point $\Delta=1$. Thus Eq.~\eqref{eq:int_charg_mom} can be approximated as the sum of two Gaussian integrals of the form
\begin{equation}
\frac{\mathrm{Tr}(\rho_{A,Q}^2)}{\mathrm{Tr}(\rho_{A}^2)}=\sum_{\alpha_0=0,\pi}A(\alpha_0)\int_{-\infty}^{\infty}\frac{\mathrm{d}\alpha}{2\pi}e^{-(\ell/\bar{\xi}_2(\Delta)+\bar{\gamma}_2(\Delta)\log\ell)(\alpha-\alpha_0)^2}.
\end{equation}
The integrals above can be straightforwardly performed. Plugging the final result in Eq.~\eqref{eq:renyiasymm}, we get the following expression for the second R\'enyi entanglement asymmetry in the large $\ell$ limit 
\begin{equation}\label{eq:u1break}
\Delta S_A^{(2)} = \frac{1}{2} \log\frac{\ell}{\bar{\xi}_2(\Delta)}-\log(\frac{A(0)+A(\pi)}{\sqrt{4\pi}})+\frac{\bar{\gamma}_2(\Delta)}{2}\frac{\bar{\xi}_2(\Delta)}{\ell}\log\ell.
\end{equation}
This result agrees with the generic formula for the R\'enyi entanglement asymmetry when a continuous symmetry breaks into a discrete subgroup (here $U(1)\to\mathbb{Z}_2$) obtained in Ref.~\cite{fadc-24} for the ground state of critical systems. 
Although the leading order behaviour, $1/2\log \ell$, derives from the exponential decay of the charged moments, which is non-universal, 
it is fixed by the dimension of the broken symmetry group, $\dim(U(1))/2\log \ell$ and $\dim(U(1))=1$. The fact that the system is critical can be understood from the presence of the $\log\ell/\ell$ term, which comes from the power-law behaviour we found in Eqs. \eqref{eq:yayfit} and \eqref{eq:resultCFTwow}. However, this term not only depends on the universal exponent $\gamma_2(\Delta)$ of the charged moments but also on the correlation length $\xi_2(\Delta)$, making it 'semi-universal'. A similar feature was found studying the $U(1)$ particle number symmetry breaking in the XY spin chain at criticality~\cite{fadc-24}. Semi-universal $\log\ell/\ell$ terms do not only appear in the entanglement asymmetry but also in other quantities such as the corner free energy of critical systems~\cite{sd-13} or the emptiness formation probability in the critical XY spin chain~\cite{stephan-14}.

\section{Symmetry breaking of a non-Abelian group}\label{sec:SU2}

In this section, we study the entanglement asymmetry corresponding to the full $SU(2)$ group of spin rotations. 

If we insert the generic expression~\eqref{eq:twirling} for $\rho_{A, G}$ in the definition of the R\'enyi entanglement asymmetry, then the latter can be obtained from the $n$-fold integral over the group
\begin{equation}\label{eq:G_group_ent_asymm}
\Delta S_A^{(n)}=\frac{1}{1-n}\log\left[\frac{1}{{\rm vol}(G)^{n-1}}\int_{G^n} {\rm d}\mathbf{g}\frac{Z_n(\mathbf{g})}{Z_n(\boldsymbol{0})}\delta\Big(\prod_{j}^ng_j\Big)\right]
\end{equation}
of the charged moments
\begin{equation}
Z_n(\mathbf{g})=\Tr(\rho_A U_{A, g_1}\dots \rho_A U_{A, g_n}).
\end{equation}
In these formulas, $G^n$ stands for the Cartesian product $G^n=G\times \stackrel{n}{\cdots} \times G$ and $\mathbf{g}$ is the $n$-tuple $\mathbf{g}=(g_1, \dots, g_n)\in G^n$.

Using the same scaling arguments as in Eq.~\eqref{eq:free_energy_riem}, we can conjecture that 
the charged moments $Z_n(\mathbf{g})$ corresponding to the $SU(2)$ group of spin rotations behave for large subsystem size $\ell$ as
\begin{equation}\label{eq:SU(2)_charged_moments}
\frac{Z_n(\mathbf{g})}{Z_n(\boldsymbol{0})}\sim e^{-T_n(\mathbf{g},\Delta)\ell}\ell^{-\beta_n(\mathbf{g}, \Delta)},
\end{equation}
where $T_n(\mathbf{g})=\sum_{j=1}^n t(g_j)$ and $t(g_j)$ is the tension of the line defect associated with the group element $g_j\in SU(2)$. The exponent $\beta_n(\mathbf{g}, \Delta)$ could be determined from the CFT, although calculating it within the perturbative approach of Sec.~\ref{sec:SU2} is very complicated.

To obtain the entanglement asymmetry, we have to plug Eq.~\eqref{eq:SU(2)_charged_moments} into the $n$-fold integral~\eqref{eq:G_group_ent_asymm} evaluated over the group $G=SU(2)$. As we have done in the previous section for the $U(1)$ subgroup, we can solve it by performing a saddle point approximation around the elements $\mathbf{h}\in SU(2)^n$ for which $Z_n(\boldsymbol{\mathbf{h}})=Z_n(\boldsymbol{0})$. These correspond to the elements of $SU(2)$ that are symmetries of the XXZ spin chain: the subgroup $U(1)$ of rotations around the $z$-axis and the $\mathbb{Z}_2$ subgroups corresponding to spin-flips around the $x$ and $y$ axes.  Since $SU(2)$ is simply connected, any group element $g\in SU(2)$ in a neighbourhood of an element $h$ of the symmetric subgroup can be written as $g=h e^{iX}$, where $X$ is an element of the $\mathfrak{su}(2)$ Lie algebra, $X\in \mathfrak{su}(2)$.
It is now useful to introduce a coordinate chart $(x_1, x_2, x_3)\in\mathbb{R}^3$ for $\mathfrak{su}(2)$, such that $g(x_1, x_2, x_3, h)=h e^{i(x_1 \sigma_x+ x_2 \sigma_y + x_3 \sigma_z)/2}$, where $\sigma_j$ are the Pauli matrices. If we rewrite the 
coefficients $T_n(\mathbf{g})$ and $\beta_n(\mathbf{g})$ in terms of the coordinates, then we can expand them around $x_1=0$ and $x_2=0$ fixing $x_3=0$. Due to the invariance under spin flips around the $z$ axis, the charged moments are even functions in the coordinates, $Z_n(\mathbf g(\mathbf x,\mathbf h))=Z_n(\mathbf g(-\mathbf x,\mathbf h))$, and 
\begin{equation}\label{eq:exp_T_beta}
\begin{aligned}
T_n(\mathbf{g}(\boldsymbol{\mathrm{x}}, \mathbf{h}))&=\frac{1}{2}\boldsymbol{\mathrm{x}}^t \mathsf{H}_{T_n}(\mathbf{h}) \boldsymbol{\mathrm{x}}+\mathcal O(\boldsymbol{\mathrm{x}^4})\\
\beta_n(\mathbf{g}(\boldsymbol{\mathrm{x}}, \mathbf{h}))&=\frac{1}{2}\boldsymbol{\mathrm{x}}^t \mathsf{H}_{\beta_n}(\mathbf{h})\boldsymbol{\mathrm{x}}+\mathcal O(\boldsymbol{\mathrm{x}^4})
\end{aligned}
\end{equation}
where $\boldsymbol{\mathrm{x}}=(\mathrm{x}_1,\dots, \mathrm{x}_n)\in\mathbb{R}^{2n}$ and $\mathrm{x}_j$ represents the $j$-copy of the coordinates $\mathrm{x}=(x_1, x_2)$.  
In Eq.~\eqref{eq:exp_T_beta}, the Hessian matrices $\mathsf{H}_{T_n}$ and $\mathsf{H}_{\beta_n}$ are formed by $n\times n$ blocks of dimension $2\times 2$. 

The composition of spin flips around the $x$ and $y$ axes, $e^{\pm i\pi \sigma_x/2}$ and $e^{\pm i\pi \sigma_y/2}$, with the rotations around the $z$ axis satisfies the identities $e^{\pm i\pi \sigma_x/2}e^{i\pi \sigma_z/2}=e^{\pm i\pi \sigma_y/2}$. Therefore, any element of $SU(2)$ that is a symmetry of the XXZ spin chain can be written as $h=e^{i\theta \sigma_z/2}p$, where $p=I$ or $e^{i\pi \sigma_x/2}$. All the saddle points of the form $(e^{i\theta_1 \sigma_z/2}p_1, \dots, e^{i\theta_n \sigma_z/2}p_n)$, with $p_j=I, e^{i\pi \sigma_x/2}$, give the same contribution. Therefore, we can approximate the $n$-fold integral~\eqref{eq:G_group_ent_asymm} over $SU(2)$ by expanding it around the elements $\mathbf{h}=(e^{i\theta_1 \sigma_z/2}, \dots, e^{i\theta_n \sigma_z/2})\in U(1)^n$ and multiplying by the number of saddle points that give the same contribution, which is $2^{n-1}$.
If we parameterise the neighbourhood of $\mathbf{h}\in U(1)^n$ by the coordinate chart $\boldsymbol{\mathrm{x}}$, we eventually have 
\begin{multline}\label{eq:saddle_point_SU(2)}
\int_{SU(2)^n}\mathrm{d}\mathbf{g}\frac{Z_n(\mathbf{g})}{Z_n(\mathbf{0})}\delta\Big(\prod_{j}g_j\Big)\\
\sim 2^{(n-1)}\mu^{n-1}\int_{U(1)^n}\mathrm{d}\mathbf{h}\,\delta\Big(\prod_j h_j\Big)\int_{\mathbb{R}^{2n}}\mathrm{d}\boldsymbol{\mathrm{x}} e^{-\frac{1}{2}\boldsymbol{\mathrm{x}}^t(\mathsf{H}_{T_n}(\mathbf{h})\ell+\mathsf{H}_{\beta_n}(\mathbf{h})\log\ell)\boldsymbol{\mathrm{x}}}\delta\Big(\sum_{j=1}^n\mathrm{x}_j\Big).
\end{multline}
In this expression, we have taken into account that the Haar measure of $SU(2)$ in the selected coordinate chart is ${\rm d}g=\mu(x_1, x_2, 0){\rm d}\mathrm{x}$ and ${\rm d}\mathbf{h}$ is the Haar measure of the $U(1)^n$ group. We have further expanded $\mu(x_1, x_2, 0)$ around $x_1=x_2=0$ and restricted to the zeroth order term $\mu:=\mu(0, 0, 0)$. The integral is extended to the whole real space due to the saddle-point approximation for large $\ell$. 
The Dirac delta over the group $\delta(\prod_j g_j)$ translates in the coordinate chart into $\delta(\prod h_j)\delta(\sum_{j=1}^n \mathrm{x}_j)/\mu$. 
In Cartesian coordinates, the Haar measure of $SU(2)$ reads~\cite{mdc-21}
\begin{equation}\label{eq:su(2)_haar_measure}
\mu(x_1, x_2, x_3)= \sqrt{2}\left(\frac{\sin(\sqrt{x_1^2+x_2^2+x_3^2}/2)}{\sqrt{x_1^2+x_2^2+x_3^2}}\right)^2,
\end{equation}
and, evaluating it at $x_1=x_2=x_3=0$, we find $\mu=1/(2\sqrt{2})$. 

Given that the coefficient $T_n(\mathbf{g})$ decomposes into the sum of the tensions of each defect line,  $T_n(\mathbf{g})=\sum_{j=1}^n t(g_j)$, the Hessian $\mathsf{H}_{T_n}$ takes a block-diagonal form, $\mathsf{H}_{T_n}=I_n\otimes \mathsf{H}_{t}$, where $\mathsf{H}_t$ is the Hessian of the single-defect tension $t(g(x_1, x_2))$, with dimension $2\times 2$. Moreover, the charged moments are invariant under the cyclic permutation of the replicas. This implies that the Hessian of $\beta_{n}(g(\boldsymbol{\mathrm{x}}))$ is a block-circulant matrix. That is, it can be written as
\begin{equation}
    \mathsf{H}_{\beta_n} = 
\begin{pmatrix}
\mathsf C_1      & \mathsf C_{n} & \cdots  & \mathsf C_3     & \mathsf C_2     \\
\mathsf C_2      & \mathsf C_1     & \mathsf C_{n} &         & \mathsf C_3     \\
\vdots   & \mathsf C_2     & \mathsf C_1     & \ddots  & \vdots  \\
\mathsf C_{n-1}  &         & \ddots  & \ddots  & \mathsf C_{n} \\
\mathsf C_{n}  & \mathsf C_{n-1} & \cdots  & \mathsf C_2     & \mathsf C_1     \\
\end{pmatrix},
\end{equation}
where the blocks $\mathsf{C}_j$ are matrices of dimension $2\times 2$.
In this case, we can employ the identity 
\begin{equation}
\int_{\mathbb{R}^{2n}}{\rm d}\boldsymbol{\mathrm{x}}
e^{-\frac{1}{2}\boldsymbol{\mathrm{x}}^t(\mathsf{H}_{T_n}\ell+\mathsf{H}_{\beta_n}\log\ell)\boldsymbol{\mathrm{x}}}\delta\Big(\sum_{j=1}^n\mathrm{x}_j\Big)=
\frac{1}{\sqrt{n}}\prod_{p=1}^{n-1}\left(\det(2\pi(\mathsf{H}_{t}(\mathbf{h})\ell+\mathsf{D}_p(\mathbf{h})\log\ell)^{-1})\right)^{1/2},
\end{equation}
found in Ref.~\cite{fadc-24}, where $\mathsf{D}_{p}$ are the Fourier transform of the blocks $\mathsf{C}_j$, $\mathsf{D}_p=\sum_{j=1}^n\mathsf{C}_j e^{\frac{i2\pi j p}{n}}$. Applying it in Eq.~\eqref{eq:saddle_point_SU(2)}, we obtain
\begin{multline}\label{eq:step1}
\int_{SU(2)^n}\mathrm{d}\mathbf{g}\frac{Z_n(\mathbf{g})}{Z_n(\mathbf{0})}\delta\Big(\prod_j g_j\Big)\\ \sim \frac{2^{(n-1)}\mu^{n-1}}{\sqrt{n}}
\int_{U(1)^n}\mathrm{d}\mathbf{h}\,\delta\Big(\prod_j h_j\Big)
\prod_{p=1}^{n-1}\left(\det(2\pi(\mathsf{H}_{t}(\mathbf{h})\ell+\mathsf{D}_p(\mathbf{h})\log\ell)^{-1})\right)^{1/2}.
\end{multline}
We can rewrite the determinant in Eq.~\eqref{eq:step1} as
\begin{equation}    \det(2\pi(\mathsf{H}_{t}(\mathbf{h})\ell+\mathsf{D}_p(\mathbf{h})\log\ell)^{-1})=\left(\frac{2\pi}{\ell}\right)^{2}\det(\mathsf{H}_{t}(\mathbf{h}))^{-1}\det(\mathbbm{1}+\mathsf{H}_{t}^{-1}(\mathbf{h})\mathsf{D}_p(\mathbf{h})\frac{\log\ell}{\ell})^{-1},
\end{equation}
and, in the limit $\ell\gg 1$, the right-hand-side above can be simplified as
\begin{equation}   \det(\mathbbm{1}+\mathsf{H}_{t}^{-1}(\mathbf{h})\mathsf{D}_p(\mathbf{h})\frac{\log\ell}{\ell})^{-1}\simeq 1-\dfrac{\log\ell}{\ell}\mathrm{Tr}(\mathsf{H}_{t}^{-1}(\mathbf{h})\mathsf{D}_p(\mathbf{h})).
\end{equation}
Plugging this result into Eq.~\eqref{eq:step1}, we get 
\begin{multline}    \int_{U(1)^n}\mathrm{d}\mathbf{h}\,\delta\Big(\prod_j h_j\Big)
\prod_{p=1}^{n-1}\left(\det(2\pi(\mathsf{H}_{t}(\mathbf{h})\ell+\mathsf{D}_p(\mathbf{h})\log\ell)^{-1})\right)^{1/2}\sim \\
\left(\frac{2\pi}{\ell}\right)^{n-1}
\int_{U(1)^n}\mathrm{d}\mathbf{h}\det(\mathsf{H}_{t}(\mathbf{h}))^{-(n-1)/2}\,\delta\Big(\prod_j h_j\Big)\left[1-\dfrac{\log\ell}{2\ell}\sum_{p=1}^{n-1}\mathrm{Tr}(\mathsf{H}_{t}^{-1}(\mathbf{h})\mathsf{D}_p(\mathbf{h}))\right].
\end{multline}
Finally, applying it in Eq.~\eqref{eq:G_group_ent_asymm} and taking into account that the volume of $SU(2)$  with respect to the Haar measure~\eqref{eq:su(2)_haar_measure}  is ${\rm vol}(SU(2))=2^{5/2}\pi^2$, the asymptotic behaviour of the entanglement asymmetry reads 
\begin{equation}\label{eq:ea_SU2}
    \Delta S_A^{(n)}=\log \ell+\log(4\pi A_n^{1/(1-n)} n^{(n-1)/2})-\dfrac{B_n}{2(1-n)A_n}\dfrac{\log\ell}{\ell},
\end{equation}
where 
\begin{equation}
A_n=\int_{U(1)^n}\mathrm{d}\mathbf{h}\det(\mathsf{H}_{t}(\mathbf{h}))^{-(n-1)/2}\,\delta\Big(\prod_j h_j\Big),
\end{equation}
and 
\begin{equation}
B_n= \int_{U(1)^n}\mathrm{d}\mathbf{h}\,\delta\Big(\prod_j h_j\Big)\det(\mathsf{H}_{t}(\mathbf{h}))^{-(n-1)/2}\sum_{p=1}^{n-1}\mathrm{Tr}(\mathsf{H}_{t}^{-1}(\mathbf{h})\mathsf{D}_p(\mathbf{h})).
\end{equation}
We can immediately recognise one crucial difference in Eq.~\eqref{eq:ea_SU2} with respect to the result found in Eq.~\eqref{eq:u1break} for the breaking of the $U(1)$ symmetry: the prefactor of the logarithmic term is $1$ rather than $1/2$. This agrees with the formula found in Ref.~\cite{cv-23} for the entanglement asymmetry in matrix product states when a continuous symmetry is broken into a continuous subgroup (here $SU(2)\to U(1)$): as in that case, the prefactor of the leading $\log \ell$ term is given by the difference between the dimensions of the two groups as $(\dim(SU(2))-\dim(U(1)))/2=1$. The novelty here is that criticality introduces a $\log\ell/\ell$ correction as in the symmetry breaking of a continuous group to a discrete subgroup analysed in Sec.~\ref{sec:asymm} and, more generally, in Ref.~\cite{fadc-24}. Again this term is semi-universal because it is a function of both the universal exponent $\beta_n(\boldsymbol{g})$ of the charged moments, which is given by the underlying CFT, and their correlation length or defect line tensions $T_n(\boldsymbol{g})$, which depends on the specific microscopic model.  

\section{Conclusions}\label{sec:concl}

In this paper, we have investigated the breaking of a $SU(2)$ symmetry 
to a $U(1)$ subgroup in a critical system through the lens of the entanglement asymmetry. As specific example, we have considered the ground state of the critical XXZ spin chain. In this system, the $SU(2)$ symmetry of spin rotations is explicitly broken except at the 
isotropic point, which corresponds to the XXX spin chain. Outside that point, only the $U(1)$ subgroup of spin rotations around the transverse direction remains as a symmetry of the system. 

We have first analysed the entanglement asymmetry associated with the broken $U(1)$ subgroup of spin rotations around the longitudinal axis. To calculate it, we have leveraged the underlying CFT description of the critical XXZ spin chain around the isotropic point: it corresponds to a self-dual compact boson, or equivalently a $SU(2)_1$ WZW theory, perturbed by a current-current term that explicitly breaks the $SU(2)$ symmetry. Applying the replica trick, the entanglement asymmetry can be related to 
the charged moments of the reduced density matrix. Within the path integral approach, the $n$ charged moment is identified with the partition function of the theory on a $n$-sheet Riemann surface with defect lines associated with different elements of the broken $U(1)$ subgroup plugged along the surface branch cuts. We have seen that this partition function can be alternatively written as a $2n$-point correlation function of vertex operators inserted at the branch points of the Riemann surface. 
By applying conformal perturbation theory, uniformising the Riemann surface and regularising the UV divergences, we have obtained an asymptotic expression for the $n=2$ charged moments at second order perturbation theory. We have thoroughly checked this result against numerical calculations on the lattice using tensor network techniques. In particular, we have found that on the lattice, apart from the universal algebraic term predicted by the CFT, the charged moments present an extra dominant factor that decays exponentially with the subsystem size. This term depends on the sum of the tensions of the defect lines; therefore, it is non-universal and cannot be captured by the CFT. Unfortunately, we lack the technology on the lattice to analytically determine these tensions. From these results, we have derived the asymptotic expression of the entanglement asymmetry for the broken $U(1)$ subgroup. We have found that, at leading order, it grows with the subsystem length $\ell$ as $1/2\log\ell$, where the coefficient is half of the dimension of the $U(1)$ subgroup. Criticality gives rise to a $\log\ell/\ell$ correction, with a coefficient that is semi-universal as it depends on the defect tensions. This result is in agreement with the generic expression obtained for a compact Lie group in Ref.~\cite{fadc-24} in critical systems.  

We have also considered the entanglement asymmetry of the full $SU(2)$ group of spin rotations. In this case, the corresponding charged moments are the expectation value of non-local operators
on the Riemann surface, which are difficult to calculate within the perturbed CFT approach that we have employed for the $U(1)$ subgroup. However, applying generic scaling arguments, we have deduced a functional asymptotic expression for the charged moments, from which we have obtained the entanglement asymmetry of the $SU(2)$ group. A crucial point in this case is that there is a $U(1)$ subgroup that is 
a symmetry of the XXZ spin chain. This implies that the entanglement asymmetry grows as $\log\ell$, where the coefficient is given by the difference between the dimensions of the $SU(2)$ group and the $U(1)$ subgroup that remains a symmetry of the system. Criticality yields a semi-universal $\log\ell/\ell$ correction. 

We leave several open problems. We have obtained explicit expressions for the entanglement asymmetry in terms of the universal exponent of the charged moments and the defect tensions. However, both the universal exponent and the defect tensions remain unknown for the $SU(2)$ group, as well as for the broken $U(1)$ subgroup in the case of the tensions. Moreover, the analytic study carried out here relies on the perturbed CFT description of the XXZ spin chain around the isotropic point, but it would be desirable to extend it to the full critical line. Finally, the $SU(2)$-broken state analysed here could be an interesting starting point to study a global quantum quench: by evolving the system with a Hamiltonian which restores the $SU(2)$ symmetry or its subgroup, one could wonder whether interesting phenomena like the quantum Mpemba effect mentioned in the introduction still occur.
\newline

\noindent\textbf{Acknowledgments}
\newline 

\noindent We would like to thank Mario Collura and Stefan Groha for sharing with us the detailed perturbative calculation and the numerical data that appear in Ref.~\cite{collura-essler-groha}. PC and FA acknowledge support from ERC under Consolidator Grant number 771536 (NEMO). SM thanks the support from the Caltech Institute for
Quantum Information and Matter and the Walter Burke Institute for Theoretical Physics
at Caltech. ML thanks the support from the Munich Center for Quantum Science and Technology and the TUM School of Natural Sciences.

\appendix

\section{Details on the conformal perturbative calculation of the charged moments}
In this Appendix, we derive the asymptotic behaviour for large subsystem size $\ell$ of the integrals in Eqs.~\eqref{eq:fointegral},~\eqref{eq:int_pp} and~\eqref{eq:int_pm} that enter in the calculation of the charged moments using the perturbed CFT~\eqref{eq:action_pert}.

\subsection{First order}\label{app:fo_int}

Let us first consider the integral~\eqref{eq:fointegral} and omit the prefactors,
\begin{equation}\label{eq:foint_no_pref}
\tilde{I}_1=\int_{R_{\ell/b}}\mathrm{d}^2 z \left|\sum_{j=1}^n\beta_j\left(\frac{1}{z-u'_j}+\frac{1}{v'_j-z}\right)\right|^2,
\end{equation}
where $R_{\ell/b}=\{z\in \mathbb{C},\,\sqrt[n]{b/\ell}<|z|<\sqrt[n]{\ell/b}\}$. By directly performing a series expansion in $\epsilon$ of the integrand, we obtain at leading order in $\ell$ that
\begin{equation}
    \tilde I_1\approx\int_{R_{\ell/b}}\dd^2z \left|\sum_{j=1}^n\frac{\beta_j}{z}\right|^2=\frac{4\pi}{n}\left|\sum_{j=1}^n\beta_j\right|^2\log\left(\frac{\ell}{b}\right).
\end{equation}
A more careful approach based on the residue theorem provides the same result. If we separate the holomorphic and antiholomorphic parts of the integrand in Eq.~\eqref{eq:foint_no_pref} and we take polar coordinates $z=r e^{i\phi}$, then
\begin{equation}
    \tilde I_1=\int_{\sqrt[n]{b/\ell}}^{\sqrt[n]{\ell/b}} r\,\dd r\,\int_{0}^{2\pi}\dd\phi \sum_{j, k=1}^n\beta_j\beta_{k}\left(\frac{1}{re^{i\phi}-u_j'}+\frac{1}{v_j'-re^{i\phi}}\right)\left(\frac{1}{re^{-i\phi}-u_{k}'^*}+\frac{1}{v_{k}'^*-re^{-i\phi}}\right).
\end{equation}
Performing the change of variables $w=e^{i\phi}$ ($\dd w = ie^{i\phi}\,\dd\phi$), 
the integral on the angle $\phi$ becomes a contour integral on the complex plane $w$ along the unit circle $\Gamma=\{w=e^{2\pi i t}, t\in[0,1]\}$,
\begin{equation}
    \tilde I_1 =-i\int_{\sqrt[n]{b/\ell}}^{\sqrt[n]{\ell/b}} r\,\dd r\,\oint_{\Gamma}\dd w \sum_{j, k=1}^n\beta_j\beta_k\left(\frac{1}{rw-u_j'}+\frac{1}{v_j'-rw}\right)\left(\frac{1}{r-wu_k'^*}+\frac{1}{wv_k'^*-r}\right).
\end{equation}
This contour integral can be computed exactly by applying the residue 
theorem,
\begin{equation}\label{eq:foint_res}
    \tilde I_1=2\pi \int_{\sqrt[n]{b/\ell}}^{\sqrt[n]{\ell/b}} r\,\dd r\sum_{j, k=1}^n\beta_j\beta_k\,\mathrm{Res}(f_{jk}, \Gamma),
\end{equation}
where $\mathrm{Res}(f_{jk}, \Gamma)$ is the total residue within the closed curve $\Gamma$ of the function
\begin{align}
    f_{jk}(w)=\left(\frac{1}{rw-u_i'}+\frac{1}{v_i'-rw}\right)\left(\frac{1}{r-wu_k'^*}+\frac{1}{wv_k'^*-r}\right).
\end{align}
This function has four poles and two of them, 
$w_{1}=u_j'/r$ ,$w_{2}=r/v_k'^*$, lie within the unit circle.
The total residue is, therefore,
\small
\begin{align}
    \mathrm{Res}(f_{jk}, \Gamma)&=\lim_{w\rightarrow w_1} (w-w_1)f_{jk}(w)+\lim_{w\rightarrow w_2} (w-w_2)f_{jk}(w)\\
    %
  %  &= \left(\frac{1}{r}+0\right)\!\left(\frac{1}{r-u_iu_j^*/r}+\frac{1}{v_j^*u_i/r-r}\right)\!+\!\left(\frac{1}{r^2/v_j^*-u_i}+\frac{1}{v_i-r^2/v_j^*}\right)\!\left(0+\frac{1}{v_j^*}\right)\\
    %
    &= \frac{1}{r^2-u_j'u_k'^*}+\frac{1}{v_k'^*u_j-r^2}+\frac{1}{r^2-v_k'^*u_j'}+\frac{1}{v_j'v_k'^*-r^2}%\approx\frac{1}{r^2}.
\end{align}
\normalsize
In the $\epsilon\rightarrow 0$ limit, 
\begin{equation}
\mathrm{Res}(f_{jk}, \Gamma)\sim \frac{1}{r^2}.
\end{equation}
Inserting this result in Eq.~\eqref{eq:foint_res}, we find
\begin{align}
    \tilde I_1=2\pi \int_{\sqrt[n]{b/\ell}}^{\sqrt[n]{\ell/b}} \frac{\dd r}{r}\sum_{j, k=1}^n\beta_j\beta_k=\frac{4\pi}{n}\left|\sum_{j=1}^n\beta_j\right|^2\log\left(\frac{\ell}{b}\right).
\end{align}

\subsection{Second order}\label{app:so_int}

Let us first consider the integral $P_+$ in the case $n=2$. According to Eq.~\eqref{eq:vu}, $u_1'=-u_2'=(\epsilon/\ell)^{1/2}\equiv u$ and $v_1'=-v_2'=(\ell/\epsilon)^{1/2}\equiv 1/u$. We split the integrand in the
holomorphic and anti-holomorphic parts,
\begin{multline}
P_+=\int_{\mathbb{C}} \frac{{\rm d}^2z {\rm d}^2 w}{(z-w)^2(\bar{z}-\bar{w})^2}\left[\left(\frac{(z-u)(z+1/u)(w+u)(w-1/u)}{(z+u)(z-1/u)(w-u)(w+1/u)}\right)^{\frac{\alpha}{2\pi}}\right.\\
\times
\left.\left(\frac{(\bar{z}-u)(\bar{z}+1/u)(\bar{w}+u)(\bar{w}-1/u)}{(\bar{z}+u)(\bar{z}-1/u)(\bar{w}-u)(\bar{w}+1/u)}\right)^{\frac{\alpha}{2\pi}}-1\right],
\end{multline}
and perform a large $\ell$ (small $u$) expansion
\begin{multline}\label{eq:exp_P_+}
P_+=\int_{\mathbb{C}}{\rm d}^2z {\rm d}^2 w\left[\left(\frac{1}{(z-w)(\bar{z}-\bar{w})^2}\left(1+\frac{1}{zw}\right)+{\rm c.c.}\right)\frac{\alpha u}{\pi}\right.\\
+\left.
\left(\frac{1}{w-z}\left(1+\frac{1}{\bar{z}\bar{w}}\right)+{\rm c.c.}\right)^2\frac{\alpha^2 u^2}{2\pi^2}+\mathcal{O}(u^3)\right],
\end{multline}
where c.c. stands for complex conjugate. The integrals above diverge. To compute them, we change the variables
to
\begin{equation}\label{eq:cm_cv}
z_{M}=\frac{z+w}{2},\quad z_m=z-w
\end{equation}
and we take as cut-offs $\sqrt{b/\ell}\leq |z_M|\leq \sqrt{\ell/b}$ and $b/\ell\leq |z_m|\leq \ell/b$. If we further consider polar coordinates, $z_M=r_M e^{i\varphi_M}$ and $z_m=r_m e^{i\varphi_m}$ with
$\varphi_M, \varphi_m\in[0, 2\pi)$, then it is straightforward to check that all the integrals in the order $\mathcal{O}(u)$ term in Eq.~\eqref{eq:exp_P_+} vanish while, at order $\mathcal{O}(u^2)$, the only non-zero integral is
\begin{equation}
\int {\rm d}^2z{\rm d}^2 w \frac{1}{|w-z|^2}\left|1+\frac{1}{\bar{z}\bar{w}}\right|^2=\frac{\pi^2}{4}\frac{\ell}{b}\log\left(\frac{\ell}{b}\right)\left(1+\mathcal{O}(\ell^{-1})\right).
\end{equation}
Inserting this result in Eq.~\eqref{eq:exp_P_+} and taking into account that $u=(\epsilon/\ell)^{1/2}$, we finally find that
\begin{equation}
P_+\simeq \frac{\alpha^2}{4}\frac{\epsilon}{b}\log\left(\frac{\ell}{b}\right),
\end{equation}
at leading order in $\ell$.

We can follow similar steps to compute the integral $P_-$ when $n=2$.
After separating the integral in the holomorphic and anti-holomorphic
parts and performing an expansion around $u=0$, we find
\begin{multline}
P_-=\int_{\mathbb{C}}{\rm d}^2z {\rm d}^2w\left[\left(\frac{1}{(z-w)(\bar{z}-\bar{w})^2}\left(1+\frac{1}{zw}\right)-{\rm c.c.}\right)\frac{\alpha u}{\pi}\right.\\
+\left.
\left(\frac{1}{w-z}\left(1+\frac{1}{\bar{z}\bar{w}}\right)-{\rm c.c.}\right)^2\frac{\alpha^2 u^2}{2\pi^2}+\mathcal{O}(u^3)\right].
\end{multline}
This expansion is exactly the same as the one in Eq.~\eqref{eq:exp_P_+}, except for the sign of the complex conjugate part of the coefficients. Therefore, we can apply the same steps as in $P_+$ and we eventually obtain
\begin{equation}
P_-\simeq -\frac{\alpha^2}{4}\frac{\epsilon}{b}\log\left(\frac{\ell}{b}\right),
\end{equation}
at leading order in $\ell$.

As for the integral $P_3'$ in Eq.~\eqref{eq:P3prime}, let us write it down without the prefactors,
\begin{equation}
    \tilde P_3' = \int_{\mathbb{C}}\frac{\mathrm{d}^2z\mathrm{d}^2w}{|z-w|^4}\mathrm{Re}\left\{(z-w)^2\left[\sum_{j=1}^n\beta_j\left(\frac{1}{z-u'_j}+\frac{1}{v_j'-z}\right)\right]\left[\sum_{j=1}^n\beta_j\left(\frac{1}{w-u'_j}+\frac{1}{v_j'-w}\right)\right]\right\}.
\end{equation}
In the case $n=2$, if we perform a large $\ell$ (small $u$) expansion of the integrand and cancel inside $\mathrm{Re}\{\cdot\}$ all the polynomial terms that are not invariant under the change $z,w\mapsto ze^{i\phi},we^{i\phi}$, which give no contribution due to the symmetries of the integral, then we find
\begin{multline}
    \tilde P_3'=4\beta^2\int_{\mathbb{C}}\frac{\mathrm{d}^2z\mathrm{d}^2w}{|z-w|^4}\mathrm{Re}\left\{\frac{(z-w)^2}{zw}\right\}+8\beta^2u^2\int_{\mathbb{C}}\frac{\mathrm{d}^2z\mathrm{d}^2w}{|z-w|^4}\mathrm{Re}\left\{\frac{(z-w)^2}{zw}\right\}+\mathcal O(u^4).
\end{multline}
These two integrals are identical and were shown to be zero in the calculation above of $P_+$ by applying the change of variables~\eqref{eq:cm_cv}. In fact, their integrand is of the form $\frac{1}{(z-w)^2\bar z\bar w}+\mathrm{c.c.}$, which is one of the contributions to the coefficient of the order $\mathcal{O}(u^2)$ term in Eq.~\eqref{eq:exp_P_+} that vanishes.

\section{Numerical techniques for the evaluation of the charged moments}\label{app:numerics}

All the numerical data in this work have been obtained employing tensor network methods; in particular, we use the TeNPy Library (version 0.10.0)~\cite{tenpy} for performing iDMRG simulations of the ground state of the XXZ Hamiltonian~\eqref{eq:xxzham}. The simulations have been performed on an infinite charge-conserving MPS in the sector of zero total magnetisation in the $z$-axis, with a maximum value $\chi=150$ for the bond dimension. Due to limitations of the iDMRG algorithm, we selected a two-site unit cell for the infinite MPS; for the XXZ model, this inevitably results in a dimerised state $\ket{\psi}$ that is not translationally invariant. If we call $T$ the unitary translation operator that maps the site $j$ into $j+1$, we can obtain the translationally-invariant ground state by constructing the state
\begin{equation}
    \ket{\psi^*} = \frac{\ket{\psi}+T\ket{\psi}}{\sqrt{2}}.
\end{equation}
From it, we are able to recover the local density matrix of the translationally-invariant ground state as:
\begin{equation}
    \rho_A^*=\frac{\rho_A+\rho_A'}{2},\qquad \rho_A=\tr_B(\ketbra{\psi}{\psi}),\quad \rho_A'=\tr_B(T\ketbra{\psi}{\psi}T^\dagger).\label{eq:mps-state}
\end{equation}
This equality is true because the mixed term $\tr_B(T\ketbra{\psi}{\psi})$ vanishes for an infinite MPS if $A$ has finite length and $\ket\psi$ is not translationally invariant. Indeed, within this approach, we are able to replicate the numerical results for the full counting statistics from Ref. \cite{collura-essler-groha}, which would otherwise display strong un-physical even-odd effects when only considering the $\ket\psi$ state resulting from the iDMRG algorithm. The full counting statistics can also be used as a benchmark for the quality of the simulated ground state: since the full counting statistics associated to the $Q_A=\sum_{j\in A}S_j^x$ charge displays a power-law behaviour for the angle $\alpha=\pi$ \cite{collura-essler-groha} for $\ell$ even, a good simulation should provide $\tr(\rho^*_Ae^{i\pi Q_A})\propto \ell^{-1/4}$.

\begin{figure}
    \centering
    \includegraphics[width=.9\textwidth]{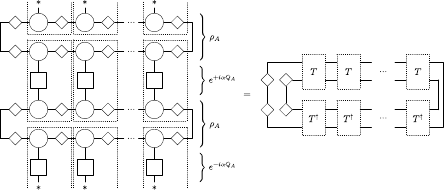}
    \caption{Diagrammatic representation of the tensor network contraction associated to the second Rényi charged moments $Z_2(\alpha)=\Tr(\rho_A e^{i\alpha Q_A} \rho_A e^{-i\alpha Q_A})$ for the case of a translationally invariant reduced density matrix $\rho_A$. Once the $T$ tensor associated to $\ell=1$ has been built, we can perform contractions efficiently by recursively evaluating $T_{2^{k+1}}=T_{2^k}\cdot T_{2^k}$ where the subscript indicates the associated system size $\ell$.}
    \label{fig:tn}
\end{figure}

Once the iMPS states have been prepared, contractions have to be performed in order to obtain the value of the charged moments for different subsystem sizes $\ell$, see Fig.~\ref{fig:tn}. Such contractions involve the construction in the computer memory of tensors of size $\chi^4$, which limit our capabilities to further increase the bond dimension for the ground state simulation. Nevertheless, thanks to the periodic structure of the infinite MPS, it is possible to reach large subsystem sizes in logarithmic time by recursively building up the necessary transfer matrices associated to subsystems of length $\ell=2,4,8,...,2^k$. Note that for the charged moments associated to the second Rényi entanglement asymmetry, we need to perform three different contractions:
\begin{multline}
    Z_2(\alpha,-\alpha)=\frac{1}{4}\left(\tr_A\left(\rho_Ae^{i\alpha S^x_A}\rho_Ae^{-i\alpha S^x_A}\right)+\tr_A\left(\rho_A'e^{i\alpha S^x_A}\rho_A'e^{-i\alpha S^x_A}\right)+\right.\\
    +\left.
    2\cdot\mathrm{Re}\left[\tr_A\left(\rho_Ae^{i\alpha S^x_A}\rho_A'e^{-i\alpha S^x_A}\right)\right]\right).
\end{multline}
The number of different contractions that need to be performed increases exponentially with the Rényi index due to the growth in the polynomial expansion of $Z_n(\boldsymbol\alpha)$ when replacing $\rho^*_A$ with Eq. \eqref{eq:mps-state}.


\begin{thebibliography}{10}

\bibitem{gross-96}
D. J. Gross,
{\it The role of symmetry in fundamental physics},
\href{https://doi.org/10.1073/pnas.93.25.14256}{PNAS {\bf 96}, 14256 (1996)}.



\bibitem{ach-22}
R. Arouca, Andrea Cappelli, T. H. Hansson,
{\it Quantum Field Theory Anomalies in Condensed Matter Physics}, 
\href{https://doi.org/10.21468/SciPostPhysLectNotes.62}{SciPost Phys. Lect. Notes 62 (2022)}.




\bibitem{laflorencie}
N. Laflorencie and S. Rachel, 
{\it Spin-resolved entanglement spectroscopy of critical spin chains
and Luttinger liquids}, 
\href{http://dx.doi.org/10.1088/1742-5468/2014/11/P11013}{J. Stat. Mech. (2014) P11013}.

\bibitem{gs-18}
M. Goldstein and E. Sela,
{\it Symmetry-resolved entanglement in many-body systems},
\href{https://doi.org/10.1103/PhysRevLett.120.200602}{Phys. Rev. Lett. {\bf 120}, 200602 (2018)}. 

\bibitem{xavier}
J. C. Xavier, F. C. Alcaraz, and G. Sierra,
{\it Equipartition of the Entanglement Entropy},
\href{https://doi.org/10.1103/PhysRevB.98.041106}{Phys. Rev. B {\bf 98}, 041106 (2018)}.

\bibitem{lukin}
A. Lukin, M. Rispoli, R. Schittko, M. E. Tai, A. M. Kaufman, S. Choi, V. Khemani, J. Leonard, and M. Greiner, 
{\it Probing entanglement in a many-body localized system}, 
\href{https://dx.doi.org/10.1126/science.aau0818}{Science {\bf 364}, 6437 (2019)}.

\bibitem{cgs-18}
E. Cornfeld, M. Goldstein, and E. Sela, 
{\it Imbalance Entanglement: Symmetry Decomposition of Negativity}, \href{https://doi.org/10.1103/PhysRevA.98.032302}{Phys. Rev. A {\bf 98}, 032302 (2018)}.

\bibitem{brc-19}
R. Bonsignori, P. Ruggiero, and P. Calabrese, {\it Symmetry resolved entanglement in free fermionic systems},
\href{https://doi.org/10.1088/1751-8121/ab4b77}{J. Phys. A  {\bf 52}, 475302 (2019)}.

\bibitem{mdgc-20}
S. Murciano, G. Di Giulio, and P. Calabrese, 
{\it Entanglement and symmetry resolution in two
dimensional free quantum field theories}, 
\href{https://link.springer.com/article/10.1007/JHEP08(2020)073}{JHEP {\bf 08} (2020) 073}.

\bibitem{crc-20}
L. Capizzi, P. Ruggiero, and P. Calabrese, 
{\it Symmetry resolved entanglement entropy of excited
states in a CFT}, 
\href{https://doi.org/10.1088/1742-5468/ab96b6}{J. Stat. Mech. (2020) 073101}.

\bibitem{mdc-21}
S. Murciano, J. Dubail, and P. Calabrese,
{\it Symmetry-resolved entanglement entropy in Wess-Zumino-Witten models},
\href{https://doi.org/10.1007/JHEP10%282021%29067}{JHEP {\bf 10} (2021) 067}.

\bibitem{chen-21}
H.-H. Chen, 
{\it Symmetry decomposition of relative entropies in conformal field theory}, 
\href{https://doi.org/10.1007/JHEP07(2021)084}{JHEP {\bf 07} (2021) 084}.

\bibitem{cc-21}
L. Capizzi and P. Calabrese, 
{\it Symmetry resolved relative entropies and distances in conformal
field theory}, 
\href{https://doi.org/10.1007/JHEP10%282021%29195}{JHEP {\bf 10} (2021) 195}.

\bibitem{mbc-21}
S. Murciano, R. Bonsignori, and P. Calabrese, 
{\it Symmetry decomposition of negativity of massless free fermions}, 
\href{https://doi.org/10.21468/SciPostPhys.10.5.111}{SciPost Phys. {\bf 10}, 111 (2021)}.

\bibitem{eimd-21}
B. Estienne, Y. Ikhlef, and A. Morin-Duchesne, 
{\it Finite-size corrections in critical symmetry-resolved entanglement}, 
\href{https://scipost.org/10.21468/SciPostPhys.10.3.054}{SciPost Phys. {\bf 10}, 054 (2021)}.

\bibitem{mt-22}
A. Milekhin and A. Tajdini, {\it Charge fluctuation entropy of Hawking radiation: a replica-free way to find large entropy},
\href{https://doi.org/10.21468/SciPostPhys.14.6.172}{SciPost Phys. {\bf 14}, 172 (2023)}.


\bibitem{acdgm-22}
F. Ares, P. Calabrese, G. Di Giulio, and S. Murciano, 
{\it Multi-charged moments of two intervals in conformal field theory}, 
\href{https://doi.org/10.1007/JHEP09%282022%29051}{JHEP {\bf 09} (2022) 051}.

\bibitem{chen-22}
H.-H. Chen, 
{\it Charged R\'enyi negativity of massless free bosons}, 
\href{https://doi.org/10.1007/JHEP02(2022)117}{JHEP {\bf 02} (2022) 117}.

\bibitem{hc-20}
D. X. Horvath and P. Calabrese, {\it Symmetry resolved entanglement in integrable field theories via form factor bootstrap},
\href{https://doi.org/10.1007/JHEP11(2020)131}{JHEP 11 (2020) 131}. 

\bibitem{ghasemi-23}
M. Ghasemi, 
{\it Universal Thermal Corrections to Symmetry-Resolved Entanglement Entropy and Full Counting Statistics}, 
\href{https://doi.org/10.1007/JHEP05(2023)209}{JHEP {\bf 05} (2023) 209}.

\bibitem{mrc-20}
S. Murciano, P. Ruggiero, and P. Calabrese, {\it Symmetry resolved entanglement in two-dimensional systems via dimensional reduction},
\href{https://doi.org/10.1088/1742-5468/aba1e5}{J. Stat. Mech. (2020) 083102}.

\bibitem{fac-23}
M. Fossati, F. Ares, and P. Calabrese, 
{\it Symmetry-resolved entanglement in critical non-Hermitian
systems}, 
\href{https://doi.org/10.1103/PhysRevB.107.205153}{Phys. Rev. B {\bf 107}, 205153 (2023)}.

\bibitem{gy-23}
H. Gaur and U. A. Yajnik, 
{\it Charge imbalance resolved R\'enyi negativity for free compact boson: Two disjoint interval case}, 
\href{https://doi.org/10.1007/JHEP02%282023%29118}{JHEP {\bf 02} (2023) 118}.

\bibitem{dgmnsz-23}
G. Di Giulio, R. Meyer, C. Northe, H. Scheppach, and S. Zhao, 
{\it On the Boundary Conformal Field Theory Approach to Symmetry-Resolved Entanglement}, 
\href{https://scipost.org/SciPostPhysCore.6.3.049}{SciPost Phys. Core {\bf 6}, 049 (2023)}.

\bibitem{fmc-23}
A. Foligno, S. Murciano, and P. Calabrese, 
{\it Entanglement resolution of free Dirac fermions on a torus}, 
\href{http://dx.doi.org/10.1007/JHEP03(2023)096}{JHEP {\bf 03} (2023) 096}.

\bibitem{northe-23}
C. Northe, 
{\it Entanglement Resolution with Respect to Conformal Symmetry}, \href{https://journals.aps.org/prl/abstract/10.1103/PhysRevLett.131.151601}{Phys. Rev. Lett. {\bf 131}, 151601 (2023)}.

\bibitem{kmop-23}
Y. Kusuki, S. Murciano, H. Ooguri, and S. Pal, 
{\it Symmetry-resolved Entanglement Entropy, Spectra \& Boundary Conformal Field Theory}, 
\href{https://link.springer.com/article/10.1007/JHEP11(2023)216}{JHEP {\bf 11} (2023) 216}.

\bibitem{mdc-24}
S. Murciano, J. Dubail, and P. Calabrese, {\it More on symmetry resolved operator entanglement,} \href{https://doi.org/10.1088/1751-8121/ad30d1}{	J. Phys. A: Math. Theor. {\bf 57} 145002 (2024).}

\bibitem{rach-24}
F. Rottoli, F. Ares, P. Calabrese, and D. X. Horvath,
{\it Entanglement entropy along a massless renormalisation flow: the tricritical to critical Ising crossover},
\href{https://doi.org/10.1007/JHEP02%282024%29053}{JHEP {\bf 02} (2024) 053}.

\bibitem{bamc-24}
A. Bruno, F. Ares, S. Murciano, and P. Calabrese, 
{\it Symmetry resolution of the computable cross-norm negativity of two disjoint intervals in the massless Dirac field theory}, \href{https://doi.org/10.1007/JHEP02%282024%29009}{JHEP {\bf 02} (2024) 009}.

\bibitem{cmc-23}
L. Capizzi, S. Murciano and P. Calabrese, {\it Full counting statistics and symmetry resolved entanglement for free conformal theories with interface defects}, \href{https://iopscience.iop.org/article/10.1088/1742-5468/ace3b8}{J. Stat. Mech. (2023) 073102.}

\bibitem{cass-24}
O. Castro-Alvaredo and L. Santamaria-Sanz,
{\it Symmetry-resolved measures in quantum field theory: A short review},
\href{https://doi.org/10.1142/S0217984924300023}{Mod. Phys. Lett. B (2024) 2430002}.

\bibitem{vek-21}
V. Vitale, A. Elben, R. Kueng, A. Neven, J. Carrasco, B. Kraus, P. Zoller, P. Calabrese, B. Vermersch, and M. Dalmonte, 
{\it Symmetry-resolved dynamical purification in synthetic quantum matter}, 
\href{https://doi.org/10.21468/SciPostPhys.12.3.106}{SciPost Phys. {\bf 12}, 106 (2022)}.

\bibitem{ncv-21}
A. Neven, J. Carrasco, V. Vitale, C. Kokail, A. Elben, M. Dalmonte, P. Calabrese, P. Zoller, B. Vermersch, R. Kueng, and B. Kraus,
{\it Symmetry-resolved entanglement detection using partial transpose moments},
\href{https://doi.org/10.1038/s41534-021-00487-y}{npj Quantum Inf. {\bf 7}, 152 (2021)}.

\bibitem{rvm-22}
A. Rath, V. Vitale, S. Murciano, M. Votto, J. Dubail, R. Kueng, C. Branciard, P. Calabrese, and B. Vermersch, 
{\it Entanglement barrier and its symmetry resolution: theory and experiment}, 
\href{https://doi.org/10.1103/PRXQuantum.4.010318}{PRX Quantum {\bf 4}, 010318 (2023)}.

\bibitem{amc-23}
F. Ares, S. Murciano, and P. Calabrese, 
{\it Entanglement asymmetry as a probe of symmetry breaking}, 
\href{https://doi.org/10.1038/s41467-023-37747-8}{Nature Communications {\bf 14}, 2036 (2023)}.

\bibitem{makc-24}
S. Murciano, F. Ares, I. Klich, and P. Calabrese,
{\it Entanglement asymmetry and quantum Mpemba effect in the
XY spin chain},
\href{https://doi.org/10.1088/1742-5468/ad17b4}{J. Stat. Mech. (2024) 013103}.

\bibitem{carc-24}
K. Chalas, F. Ares, C. Rylands, and P. Calabrese,
{\it  Multiple crossing during dynamical symmetry restoration and implications for the quantum Mpemba effect},
\href{https://doi.org/10.48550/arXiv.2405.04436}{arXiv:2405.04436}.

\bibitem{amvc-23}
F. Ares, S. Murciano, E. Vernier, and P. Calabrese,
{\it Lack of symmetry restoration after a quantum quench: an entanglement asymmetry study},
\href{https://doi.org/10.21468/SciPostPhys.15.3.089}{SciPost Phys. {\bf 15}, 089 (2023)}.

\bibitem{rkacmb-23}
C. Rylands, K. Klobas, F. Ares, P. Calabrese, S. Murciano, and B. Bertini, 
{\it Microscopic origin of the quantum Mpemba effect in integrable systems}, 
\href{https://doi.org/10.1103/PhysRevLett.133.010401}{Phys. Rev. Lett. {\bf 133}, 010401 (2024)}.

\bibitem{bkccr-24}
B. Bertini, K. Klobas, M. Collura, P. Calabrese, and C. Rylands,
{\it Dynamics of charge fluctuations from asymmetric initial states}, \href{https://doi.org/10.1103/PhysRevB.109.184312}{Phys. Rev. B {\bf 109}, 184312 (2024)}.

\bibitem{lztz-24}
S. Liu, H.-K. Zhang, S. Yin, S.-X. Zhang
{\it Symmetry restoration and quantum Mpemba effect in symmetric random circuits},
\href{https://doi.org/10.48550/arXiv.2403.08459}{arXiv:2403.08459}

\bibitem{tcdl-24}
X. Turkeshi, P, Calabrese, and A. De Luca,
{\it Quantum Mpemba Effect in Random Circuits},
\href{https://doi.org/10.48550/arXiv.2405.14514}{arXiv:2405.14514}.

\bibitem{cma-24}
F. Caceffo, S. Murciano, and V. Alba,
{\it Entangled multiplets, asymmetry, and quantum Mpemba effect in dissipative systems},
\href{https://doi.org/10.1088/1742-5468/ad4537}{J. Stat. Mech. (2024) 063103}.

\bibitem{avm-24}
F. Ares, V. Vitale, and S. Murciano,
{\it The quantum Mpemba effect in free-fermionic mixed states},
\href{https://doi.org/10.48550/arXiv.2405.08913}{arXiv:2405.08913}

\bibitem{yac-24}
S. Yamashika, F. Ares, and P. Calabrese,
{\it Entanglement asymmetry and quantum Mpemba effect in two-dimensional free-fermion systems},
\href{https://doi.org/10.48550/arXiv.2403.04486}{arXiv:2403.04486}.

\bibitem{fac-23m}
F. Ferro, F. Ares, and P. Calabrese,
{\it Non-equilibrium entanglement asymmetry for discrete groups: the example of the {\rm XY} spin chain},
\href{http://dx.doi.org/10.1088/1742-5468/ad138f}{J. Stat. Mech. (2024) 023101}.

\bibitem{cm-23}
L. Capizzi and M. Mazzoni, 
{\it Entanglement asymmetry in the ordered phase of many-body systems: the Ising Field Theory},
\href{https://doi.org/10.1007/JHEP12(2023)144}{JHEP {\bf 12} (2023) 144}.

\bibitem{joshi-24}
L. Kh. Joshi, J. Franke, A. Rath, F. Ares, S. Murciano, F. Kranzl, R. Blatt, P. Zoller, B. Vermersch, P. Calabrese, C. F. Roos, and M. K. Joshi,
{\it Observing the quantum Mpemba effect in quantum simulations}
\href{https://doi.org/10.1103/PhysRevLett.133.010402}{Phys. Rev. Lett. {\bf 133}, 010402 (2024)}.

\bibitem{Khor-23}
B. J. J. Khor, D. M. Kürkçüoglu, T. J. Hobbs, G. N. Perdue, and I. Klich,
{\it Confinement and Kink Entanglement Asymmetry on a Quantum Ising Chain},
\href{https://doi.org/10.48550/arXiv.2312.08601}{arXiv:2312.08601}.

\bibitem{krb-24}
K. Klobas, C. Rylands, and B. Bertini, {\it Translation symmetry restoration under random unitary dynamics}, \href{https://arxiv.org/abs/2406.04296}{arXiv:2406.04296}.

\bibitem{ampc-23}
F. Ares, S. Murciano, L. Piroli, and P. Calabrese,
{\it An entanglement asymmetry study of black hole radiation},
\href{https://doi.org/10.48550/arXiv.2311.12683}{arXiv:2311.12683}.

\bibitem{cv-23}
L. Capizzi and V. Vitale,
{\it A universal formula for the entanglement asymmetry of matrix product states},
\href{https://doi.org/10.48550/arXiv.2310.01962}{arXiv:2310.01962}.

\bibitem{fadc-24}
M. Fossati, F. Ares, J. Dubail, and P. Calabrese,
{\it Entanglement asymmetry in CFT and its relation to non-topological defects},
\href{https://doi.org/10.1007/JHEP05(2024)059}{JHEP {\bf 05} (2024) 059}.

\bibitem{chen-23}
M. Chen and H. Chen,
{\it R\`enyi entanglement asymmetry in 1+1-dimensional conformal field theories},
\href{https://doi.org/10.1103/PhysRevD.109.065009}{Phys. Rev. D {\bf 109}, 065009 (2024)}.

\bibitem{nc}
M. A. Nielsen and I. L. Chuang, 
{\it Quantum computation and quantum information},
\href{http://dx.doi.org/10.1017/CBO9780511976667}{Cambridge University Press, Cambridge, UK, 10th anniversary ed. (2010)}.

\bibitem{bbpssw-96}
C. H. Bennett, G. Brassard, S. Popescu, B. Schumacher, J. A. Smolin, and W. K. Wootters,
{\it Purification of Noisy Entanglement and Faithful Teleportation via Noisy Channels},
\href{DOI:https://doi.org/10.1103/PhysRevLett.76.722}{Phys. Rev. Lett. {\bf 76}, 722 (1996)}.

\bibitem{bdvsw-96}
C. H. Bennett, D. P. DiVincenzo, J. A. Smolin, and W. K. Wootters,
{\it Mixed-state entanglement and quantum error correction},
\href{https://doi.org/10.1103/PhysRevA.54.3824}{Phys. Rev. A {\bf 54}, 3824 (1996)}.

\bibitem{vawj-08}
J. A. Vaccaro, F. Anselmi, H. M. Wiseman, and K. Jacobs, {\it Tradeoff between extractable mechanical work, accessible entanglement, and ability to act as a reference system, under arbitrary superselection rules}, \href{https://doi.org/10.1103/PhysRevA.77.032114}{Phys. Rev. A {\bf 77}, 032114 (2008).} 

\bibitem{gms-09}
G. Gour, I. Marvian, and R. W. Spekkens, 
{\it Measuring the quality of a quantum reference frame: The relative entropy of frameness,} \href{https://doi.org/10.1103/PhysRevA.80.012307}{Phys. Rev. A {\bf 80}, 012307 (2009).}

\bibitem{t-19}
R. Takagi, {\it Skew informations from an operational view via resource theory of asymmetry,}
\href{https://www.nature.com/articles/s41598-019-50279-w}{Scien. Rep. {\bf 9}, 14562 (2019).}

\bibitem{ms-14}
I. Marvian, R. W. Spekkens, {\it Extending Noether's theorem by quantifying the asymmetry of quantum states,}
\href{https://doi.org/10.1038/ncomms4821}{Nat. Comm. {\bf 5}, 3821 (2014).}



\bibitem{tenpy} 
J. Hauschild and F. Pollmann, 
\emph{Efficient numerical simulations with Tensor Networks: Tensor Network Python (TeNPy)}, 
\href{https://scipost.org/SciPostPhysLectNotes.5}{SciPost Phys. Lect. Notes 5 (2018)}.

\bibitem{mhms-22}
Z. Ma, C. Han, Y. Meir, and E. Sela, 
{\it Symmetric inseparability and number entanglement in charge conserving mixed states}, 
\href{https://doi.org/10.1103/PhysRevA.105.042416}{Phys. Rev. A {\bf 105}, 042416 (2022)}.

\bibitem{hms-23}
C. Han, Y. Meir, and E. Sela, 
{\it Realistic Protocol to Measure Entanglement at Finite Temperatures}, 
\href{https://doi.org/10.1103/PhysRevLett.130.136201}{Phys. Rev. Lett. {\bf 130}, 136201 (2023)}.

\bibitem{giamarchi}
T. Giamarchi, 
{\it Quantum Physics in One Dimension}, 
\href{https://doi.org/10.1093/acprof:oso/9780198525004.001.0001}{Oxford Univ. Press (2007)}.

\bibitem{gaudin}
M. Gaudin, {\it The Bethe Wavefunction}, 
\href{https://doi.org/10.1017/CBO9781107053885}{Cambridge University Press  (2014)}.

\bibitem{affleck-85}
I. Affleck,
{\it Critical behavior of two-dimensional systems with continuous symmetries},
\href{https://doi.org/10.1103/PhysRevLett.55.1355}{Phys. Rev. Lett. {\bf 55}, 1355 (1985)}.

\bibitem{lukyanov-98}
S. Lukyanov,
{\it Low energy effective Hamiltonian for the XXZ spin chain},
\href{https://doi.org/10.1016/S0550-3213%2898%2900249-1}{Nucl.Phys. B {\bf 522}, 533 (1998)}.

\bibitem{lt-03}
S. Lukyanov and V. Terras,
{\it Long-distance asymptotics of spin-spin correlation functions for the XXZ spin chain},
\href{https://doi.org/10.1016/S0550-3213%2802%2901141-0}{Nucl. Phys. B {\bf 654}, 323 (2003)}.

\bibitem{yellowbook}
P.~Di~Francesco, P.~Mathieu and D.~Senechal,
\emph{Conformal Field Theory},
\href{https://doi.org/10.1007/978-1-4612-2256-9}{Springer, New York, USA (1997)}. 

\bibitem{collura-essler-groha}
M. Collura, F. Essler, and S. Groha, \emph{Full counting statistics in the spin-1/2 Heisenberg XXZ chain}, 
\href{https://doi.org/10.1088/1751-8121/aa87dd}{J. Phys. A: Math Theor. {\bf 50}, 414002 (2017)}.

\bibitem{hlw-94}
C. Holzhey, F. Larsen, and F. Wilczek,
{\it Geometric and Renormalized Entropy in Conformal Field Theory},
\href{https://doi.org/10.1016/0550-3213%2894%2990402-2}{Nucl. Phys. B {\bf 424}, 443 (1994)}.

\bibitem{cc-04}
P. Calabrese and J. Cardy,
{\it Entanglement Entropy and Quantum Field Theory},
\href{https://doi.org/10.1088/1742-5468/2004/06/P06002}{J. Stat. Mech. (2004) P06002}.

\bibitem{lm-01}
O. Lunin and S. D. Mathur,
{\it Correlation functions for $M^N/S_N$ orbifolds},
\href{https://doi.org/10.1007/s002200100431}{Commun. Math. Phys. {\bf 219}, 399 (2001)}.

\bibitem{msdz-17}
J. Maldacena, D. Simmons-Duffin, and A. Zhiboedov,
{\it Looking for a bulk point},
\href{http://dx.doi.org/10.1007/JHEP01(2017)013}{JHEP {\bf 01} (2017) 013}.

%\bibitem{acdgm-22}
%F. Ares, P. Calabrese, G. Di Giulio, and S. Murciano,
%{\it Multi-charged moments of two intervals in conformal field theory},
%\href{https://doi.org/10.1007/JHEP09%282022%29051}{JHEP {\bf 09} (2022) 051}

\bibitem{ccdgm-20}
P. Calabrese, M. Collura, G. Di Giulio, and S. Murciano,
{\it Full counting statistics in the gapped XXZ spin chain},
\href{https://doi.org/10.1209/0295-5075/129/60007}{EPL {\bf 129}, 60007 (2020)}.

\bibitem{ccen-09}
P. Calabrese, M. Campostrini, F. Essler, and B. Nienhuis, 
{\it Parity effects in the scaling of block entanglement in gapless spin chains}, \href{http://dx.doi.org/10.1103/PhysRevLett.104.095701}{Phys. Rev. Lett. {\bf 104}, 095701 (2010)}.

\bibitem{ce-10}
P. Calabrese and F. H. L. Essler,
{\it Universal corrections to scaling for block entanglement in spin-1/2 XX chains},
\href{http://dx.doi.org/10.1088/1742-5468/2010/08/P08029}{J. Stat. Mech. P08029 (2010)}. 


\bibitem{knizhnik-87}
V. Knizhnik, 
{\it Analytic fields on Riemann surfaces. II}, 
\href{http://dx.doi.org/10.1007/BF01225373}{Comm. Math. Phys. {\bf 112}, 567 (1987)}.

\bibitem{dixon-87}
L. J. Dixon, D. Friedan, E. J. Martinec, and S. H. Shenker, 
{\it The Conformal Field Theory Of Orbifolds}, 
\href{https://doi.org/10.1016/0550-3213(87)90676-6}{Nucl. Phys. B {\bf 282}, 13 (1987)}.

\bibitem{cp-88}
J. Cardy and I. Peschel, 
{\it Finite-size dependence of the free energy in two-dimensional critical systems}, 
\href{https://doi.org/10.1016/0550-3213(88)90604-9}{Nucl. Phys. B {\bf 300}, 377 (1988)}.

\bibitem{sd-13}
J.-M. St\'ephan and J. Dubail, 
{\it Logarithmic correction to the free energy from sharp corners with
angle $2\pi$}, 
\href{https://iopscience.iop.org/article/10.1088/1742-5468/2013/09/P09002}{J. Stat. Mech. (2013) P09002}.

\bibitem{stephan-14}
J.-M. St\'ephan, 
{\it Emptiness formation probability, Toeplitz determinants, and conformal field theory}, 
\href{https://doi.org/10.1088%2F1742-5468%2F2014%2F05%2Fp05010}{J. Stat Mech. (2014) P05010}.


\end{thebibliography}
\end{document}